\documentclass[a4paper,fleqn,usenatbib]{mnras}
\usepackage{mathrsfs}
\usepackage{color}
\usepackage{mathptmx}

\usepackage[T1]{fontenc}
\usepackage{ae,aecompl}
\usepackage{mathrsfs}

\usepackage{graphicx}	
\usepackage{amsmath}	
\usepackage{amssymb}	
\usepackage{csquotes}   
\DeclareMathAlphabet{\pazocal}{OMS}{zplm}{m}{n}

\def\Lya{Ly$\rm{\alpha}$~} 
\def\kms{${\rm km\,s}^{-1}$}
\def\mpc{$h^{-1}\mathrm{cMpc}\,$}

\def\to{$T_{\rm 0}\,$}

\def\taueff{$\tau_{\rm{eff}}\,$}

\def\mpc{$h^{-1} {\rm cMpc}\,$}
\def\taueff{$\tau_{\rm eff}\,$}
\newcommand{\comment}[1]{}
\def\bhi{$b_{\rm HI}\,$}
\def\nhi{$N_{\rm HI}\,$}

\newcommand{\angstrom}{\mbox{\normalfont\AA}}
\def\HI{\hbox{H$\,\rm \scriptstyle I\ $}}

\def\HeII{\hbox{He$\,\rm \scriptstyle II\ $}}

\begin{document}
\title[The low redshift \Lya forest]{The effect of stellar and AGN
  feedback on the low redshift Lyman-$\alpha$ forest in the Sherwood
  simulation suite}

\author[F. Nasir et al.]{Fahad Nasir$^{1}$\thanks{E-mail: ppxfn@nottingham.ac.uk}, James S. Bolton$^{1}$, Matteo Viel$^{2,3,4}$, Tae-Sun Kim$^{5}$, \newauthor Martin G. Haehnelt$^{6}$, Ewald Puchwein$^{6}$ \& Debora Sijacki$^{6}$
\\$^{1}$School of Physics and Astronomy, University of Nottingham, University Park, Nottingham, NG7 2RD, UK
\\$^{2}$SISSA - International School for Advanced Studies, Via Bonomea 265, I-34136 Trieste, Italy 
\\$^{3}$INAF - Osservatorio Astronomico di Trieste, Via G.B. Tiepolo 11, I-34131 Trieste, Italy 
\\$^{4}$INFN - National Institute for Nuclear Physics, Via Valerio 2, I-34127 Trieste, Italy 
\\$^{5}$Department of Astronomy, University of Wisconsin-Madison, 475 N. Charter St., Madison, WI 53706, USA
\\$^{6}$Kavli Institute for Cosmology and Institute of Astronomy, Madingley Road, Cambridge, CB3 0HA, UK}


\label{firstpage}
\pagerange{\pageref{firstpage}--\pageref{lastpage}}
\maketitle

\begin{abstract}
We study the effect of different feedback prescriptions on the
properties of the low redshift ($z\leq1.6$) \Lya forest using a
selection of hydrodynamical simulations drawn from the Sherwood
simulation suite.  The simulations incorporate stellar feedback, AGN
feedback and a simplified scheme for efficiently modelling the low
column density \Lya forest.  We confirm a discrepancy remains between
\emph{Cosmic Origins Spectrograph} (COS) observations of the \Lya
forest column density distribution function (CDDF) at $z \simeq 0.1$
for high column density systems ($N_{\rm HI}>10^{14}\rm\,cm^{-2}$), as
well as \Lya velocity widths that are too narrow compared to the COS
data. Stellar or AGN feedback -- as currently implemented in our
simulations -- have only a small effect on the CDDF and velocity width
distribution.  We conclude that resolving the discrepancy between the
COS data and simulations requires an increase in the temperature of
overdense gas with $\Delta=4$--$40$, either through additional \HeII
photo-heating at $z>2$ or fine-tuned feedback that ejects overdense
gas into the IGM at just the right temperature for it to still
contribute significantly to the \Lya forest.  Alternatively a larger,
currently unresolved turbulent component to the line width could
resolve the discrepancy.
\end{abstract}


\section{Introduction} \label{sec:intro}

The \Lya forest is an indispensable tool for probing the evolution and
distribution of gas in the intergalactic medium \citep[see][for
  reviews]{Meiksin2009,McQuinn2016}. At intermediate redshifts ($2\la
z\la5$), the majority of the intergalactic medium (IGM) by volume is
in the form of warm ($T\sim 10^{4}\rm\,K$) diffuse gas that is kept
photo-ionised by the metagalactic ultraviolet background (UVB).  The
latest hydrodynamical simulations of the IGM that model the \Lya
forest of absorption produced by this gas are in very good agreement
with a wide range of spectroscopic data from optical, ground-based
telescopes \citep[e.g.][]{Bolton_2017MNRAS}.  However, at lower
redshifts ($z\la 2$), due to the interplay between cosmic expansion,
the growth of structure and the declining intensity of the UVB, the
\Lya forest thins out and becomes more transmissive
\citep{Danforth_2016ApJ}.  By $z\sim 0$ a significant fraction of the
baryons in the IGM are in a collisionally ionised, shock heated phase
with temperatures $T\sim 10^{5}$--$10^{7}\rm\,K$ \citep{Cen1999}. In
the present day Universe, the \Lya forest therefore traces less than
one half of the remaining diffuse gas in the IGM
\citep{Lehner_2007ApJ,Shull_2012ApJ}.

Over the past two decades, the properties of \Lya absorption arising
from the low redshift IGM have been studied with observations made by
a series of space-based UV spectrographs
\citep{Bahcall1993,Weymann_1998ApJ,Janknecht_2006A&A,Kirkman_2007MNRAS,Williger_2010MNRAS,Tilton_2012ApJ}.
Most recently, the \emph{Cosmic Origins Spectrograph}
\citep[COS,][]{Green_2012ApJ}, mounted on the \emph{Hubble Space
  Telescope} (HST) has collected many more active galactic nuclei
(AGN) spectra with improved signal-to-noise, providing new
opportunities for absorption line studies of the low redshift \Lya
forest \citep{Shull_2014ApJ,Danforth_2016ApJ,Pachat_2016MNRAS}.  Some
of the key questions addressed by recent investigations are related to
the nature of these \Lya absorbers, and their relationship to the
evolution of the UVB and galactic feedback \citep[see
  e.g.][]{Dave_2010MNRAS,Tepper-Garcia_2012MNRAS,Kollmeier_2014ApJ,Shull_2015ApJ,Khaire_2016MNRAS,Viel_2016,Gurvich_2016}.

The mostly widely used measurements of \Lya absorbers at $z<2$ are the
column density distribution function (CDDF), the velocity (Doppler)
widths of the absorption lines and the evolution of the absorption
line number density with redshift. The CDDF shape and normalisation is
determined by the density distribution of absorbers and the intensity
of the UVB. The velocity width distribution provides constraints on
the gas temperature. Furthermore, both observables are in principle
sensitive to galactic feedback. In this context, hydrodynamical
simulations have significantly aided our understanding and
interpretation of these observational data
\citep{Rauch_1997ApJ,Theuns1998_lowz,Dave_1999ApJ,Richter_2006A&A,Paschos_2009MNRAS,Oppenheimer2009,Dave_2010MNRAS,Rahmati_2013MNRAS}.

Of particular note is the recent comparison of the CDDF for \Lya
absorbers at $z\simeq 0.1$ observed with COS \citep{Danforth_2016ApJ}
to hydrodynamical simulations.  \citet{Kollmeier_2014ApJ} reported
that the UVB intensity needed to match the COS measurement of the CDDF
is a factor of five larger than predicted by empirically calibrated
UVB models \citep[][HM12]{Haardt_2012ApJ}. The claimed ``photon
underproduction crisis'' prompted further investigation into this
apparent discrepancy.  Large uncertainties in the production rate of
ionising photons by quasars and star forming galaxies, and possibly
also the role of AGN feedback, have been suggested as the likely cause
\citep{Khaire2015,Shull_2015ApJ,Wakker_2015ApJ,Gurvich_2016,Gaikwad_2016}.

\citet{Viel_2016} (hereafter V17) also recently considered this
problem using hydrodynamical simulations including Illustris
\citep{Vogelsberger2014} and a sub-set of the Sherwood simulations
\citep{Bolton_2017MNRAS}.  They concluded the HM12 photo-ionisation
rate needed to be increased by a factor of two in the hydrodynamical
simulations to match the observed CDDF, smaller than the factor of
five found by \citet{Kollmeier_2014ApJ}. However, in contrast to other
recent studies, V17 also focussed on the distribution of \Lya line
velocity widths \citep[see also][]{Gaikwad17}.  V17 found that the
simulations struggled to reproduce the \Lya velocity width
distribution at $z\simeq0.1$, producing line widths that were too
narrow compared to the COS data. This was attributed to gas in the
simulations that was either too cold to produce the observed line
widths, or too highly (collisionally) ionised to produce \Lya
absorption due to vigorous AGN feedback heating the gas to
$T>10^{5}\rm \,K$ (particularly in the Illustris simulation).  V17
concluded that the \Lya line width distribution at $z\simeq 0.1$ is a
valuable diagnostic for models of feedback in the low redshift
Universe.

The goal of the present work is to the extend the V17 analysis, and
present in full the low redshift ($0<z<2$) IGM models that V17 used
from the Sherwood simulation suite \citep{Bolton_2017MNRAS}.
Following V17, we perform the same Voigt profile analysis on the
simulations and COS data to minimise potential biases in the
comparison due to line fitting. This furthermore enables a detailed
assessment of the effect that our stellar and AGN feedback
implementation has on the \Lya forest.

The paper is organised as follows. In Section 2, we briefly describe
the simulations used in this work. In Section 3, we compare the
simulations to current observational data at $z<2$ and assess the
impact of different galactic feedback prescriptions on \Lya forest
observables, including the the \Lya velocity width distribution.  The
physical origin of the absorbers is examined in Section 4. Finally, we
summarise our results in Section 5. A series of convergence tests are
given in the Appendix. Throughout this paper we refer to
comoving distances with the prefix ``c''. The cosmological parameters
used throughout this work are $\Omega_{\rm m}=0.308$, $\Omega_{\rm
  \Lambda}=0.692$, $h=0.678$, $\Omega_{\rm b}=0.0482$, $\sigma_{\rm
  8}=0.829$ and $n=0.961$ consistent with the best fit $\Lambda {\rm
  CDM}+Planck+WP+highL+BAO$ cosmological parameters
\citep{Planck_2014A&A}.


\section{Methodology}
\subsection{Hydrodynamical simulations}\label{sec:sim}

The simulations used in this work are summarised in
Table~\ref{tab:simulation}. These were performed using
\textsc{P-Gadget-3}, a modified version of the publicly available code
\textsc{Gadget-2} \citep{Springel_2005MNRAS}. The initial conditions
were generated at $z=99$ on a Cartesian grid using the
\textsc{N-GenIC} code \citep{Springel_2005Nat} using transfer
functions generated by \textsc{CAMB} \citep{Lewis_2000ApJ}.  A
detailed description of these models may be found in the Sherwood
overview paper \citep{Bolton_2017MNRAS}.

\begin{table}
\caption{The hydrodynamical simulations used in this work. The first
  column uses the naming convention \emph{L-N-param}. Here $L$ is the
  box size in units of \mpc, $N$ is the cube root of the total number
  of gas particles and \emph{-param} is associated with the subgrid
  treatment of star formation and feedback.  Here $-ps13$ refers to
  the star formation and energy driven outflow model of
  \citet{Puchwein_2013MNRAS}, while \emph{-ps13+agn} also includes AGN
  feedback based on \citet{Sijacki2007}.  All other models instead
  convert all cold ($T<10^{5}\rm\,K$), dense ($\Delta>1000$) gas into
  collisionless particles.  The remaining columns give the mass of
  each dark matter and gas particle in $h^{-1} M_{\sun}$ and the
  gravitational softening length in $h^{-1} \rm{ckpc}$. All
  simulations are performed to $z=0$.}
\begin{center}
 \begin{tabular}{|c|c|c|c|}

  \hline
  Name           & $M_{\rm dm} $       & $M_{\rm gas}$      & $l_{\rm soft}$    \\
  & $[h^{-1} M_{\sun}]$ & $[h^{-1} M_{\sun}]$ & $[h^{-1} \rm{ckpc}]$ \\
  \hline
      80-512-ps13 & $2.75\times 10^8$ & $5.10\times 10^7$ & $6.25$ \\   
      80-512-ps13+agn& $2.75\times 10^8$ & $5.10\times 10^7$ & $6.25$ \\   
      80-512         & $2.75\times 10^8$ & $5.10\times 10^7$ & $6.25$      \\  
      80-1024        & $3.44\times 10^7$ & $6.38\times 10^6$ & $3.13$    \\
      
      80-256         & $2.20\times 10^9$ & $4.08\times 10^8$ & $12.50$     \\   
      40-512         & $3.44\times 10^7$ & $6.38\times 10^6$ & $3.13$       \\   
      20-256         & $3.44\times 10^7$ & $6.38\times 10^6$ & $3.13$     \\   
      \hline

 \end{tabular}
\label{tab:simulation}
\end{center}
\end{table}

The photo-ionisation and photo-heating rates are calculated using a
spatially uniform UVB model for emission from star forming galaxies
and quasars \citep{Haardt_2012ApJ} applied in the optically thin
limit. A small increase to the \HeII photo-heating rate,
$\epsilon_{\rm HeII}=1.7\epsilon_{\rm HeII}^{\rm HM12}$ at $2.2<z<3.4$
was applied to better match observational measurements of the IGM
temperature at $z>2$ \citep{Becker_2011MNRAS}.  The ionised fractions
for hydrogen and helium are obtained assuming ionisation equilibrium.
Metal line cooling is not included, although this is not expected to
impact significantly on the \Lya forest at the redshifts considered
here \citep{Tepper-Garcia_2012MNRAS}.  In five of the simulations, gas
particles with temperatures $T<10^5$K and densities
$\Delta=\rho/\langle \rho \rangle >1000$ are converted into
collisionless star particles \citep[the \textsc{QUICKLYA} method,
  see][]{Viel2004} to speed up the calculation. We will show this
choice makes very little difference to \Lya forest systems at low
column densities, $N_{\rm HI}<10^{14.5}\rm\,cm^{-2}$, in our analysis.
However, galactic winds and AGN feedback may have an impact on high
column density lines that probe the (circumgalactic) gas in the
vicinity of dark matter haloes. Therefore, two models that follow star
formation and feedback are included. The first model (80-512-ps13)
includes energy driven galactic outflows following
\citet{Puchwein_2013MNRAS}.  The star formation model is based on
\citet{Springel_2003MNRAS}, but assumes a Chabrier rather than
Salpeter initial mass function (IMF) and a galactic wind velocity that
is directly proportional to the escape velocity of the galaxy.  These
choices increase the available supernovae feedback energy by a factor
of two, as well as raising the mass-loading of winds in low-mass
galaxies.  The second run (80-512-ps13+agn) also incorporates AGN
feedback, again following \citet{Puchwein_2013MNRAS}. This AGN
feedback model is based on \citet{Sijacki2007}. In the ``quasar''
mode, when accretion rates are above $0.01$ of the Eddington rate,
only $0.5$ per cent of the accreted rest mass energy is thermally
coupled to the surrounding gas. For lower accretion rates a more
efficient coupling of AGN jets in the ``radio'' mode is assumed and
$2$ per cent of the rest mass energy is used for recurrently injecting
hot AGN bubbles.

\begin{figure*}
    \centering
    \begin{minipage}{.495\textwidth}
        \centering
        \includegraphics[trim={1.0cm 0.0cm 0.0cm 1.0cm}, clip=true, width=\columnwidth]{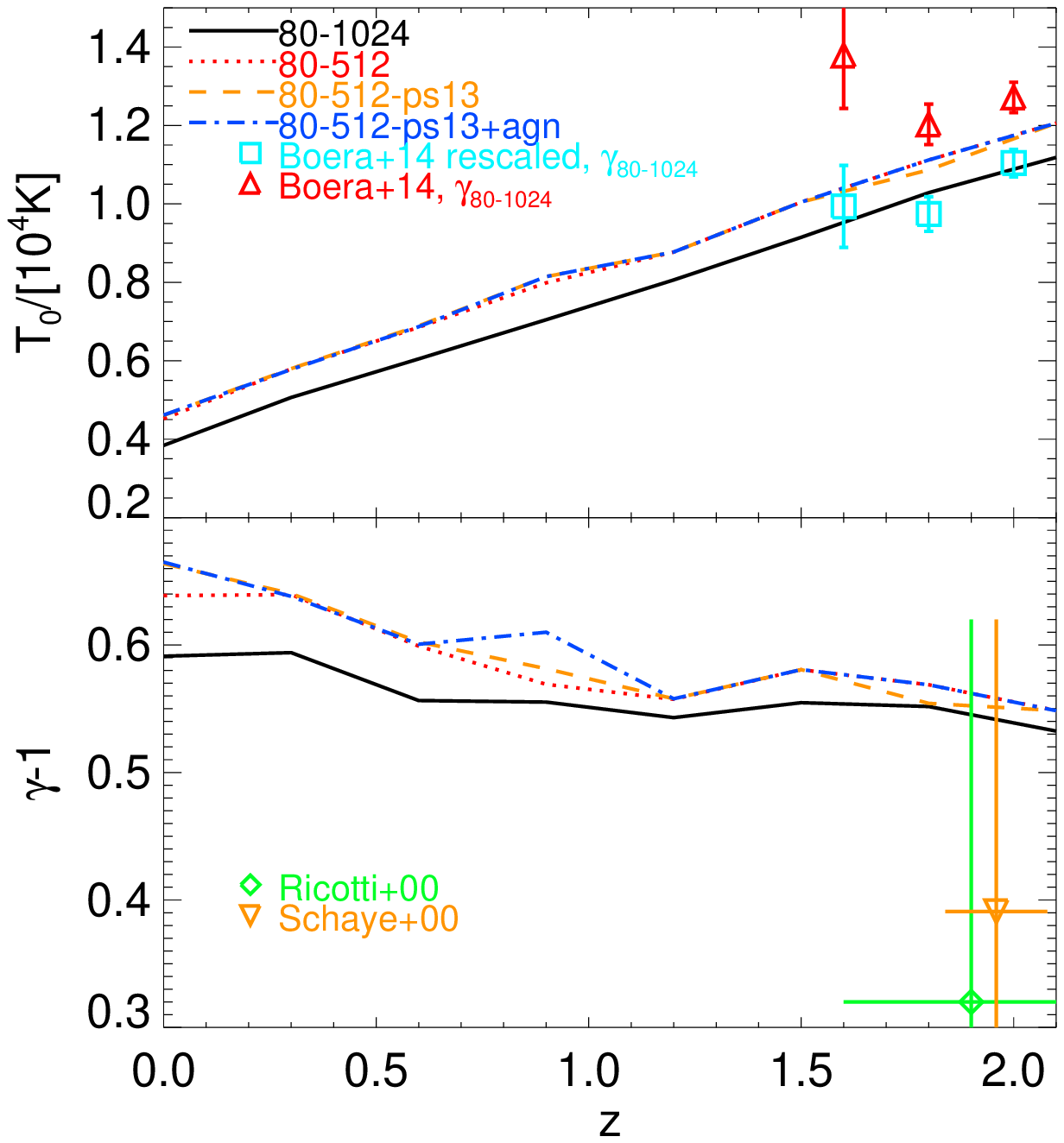}       
    \end{minipage}
    \begin{minipage}{.495\textwidth}
        \centering
        \includegraphics[trim={1.0cm 0.0cm 0.0cm 1.0cm}, clip=true, width=\columnwidth]{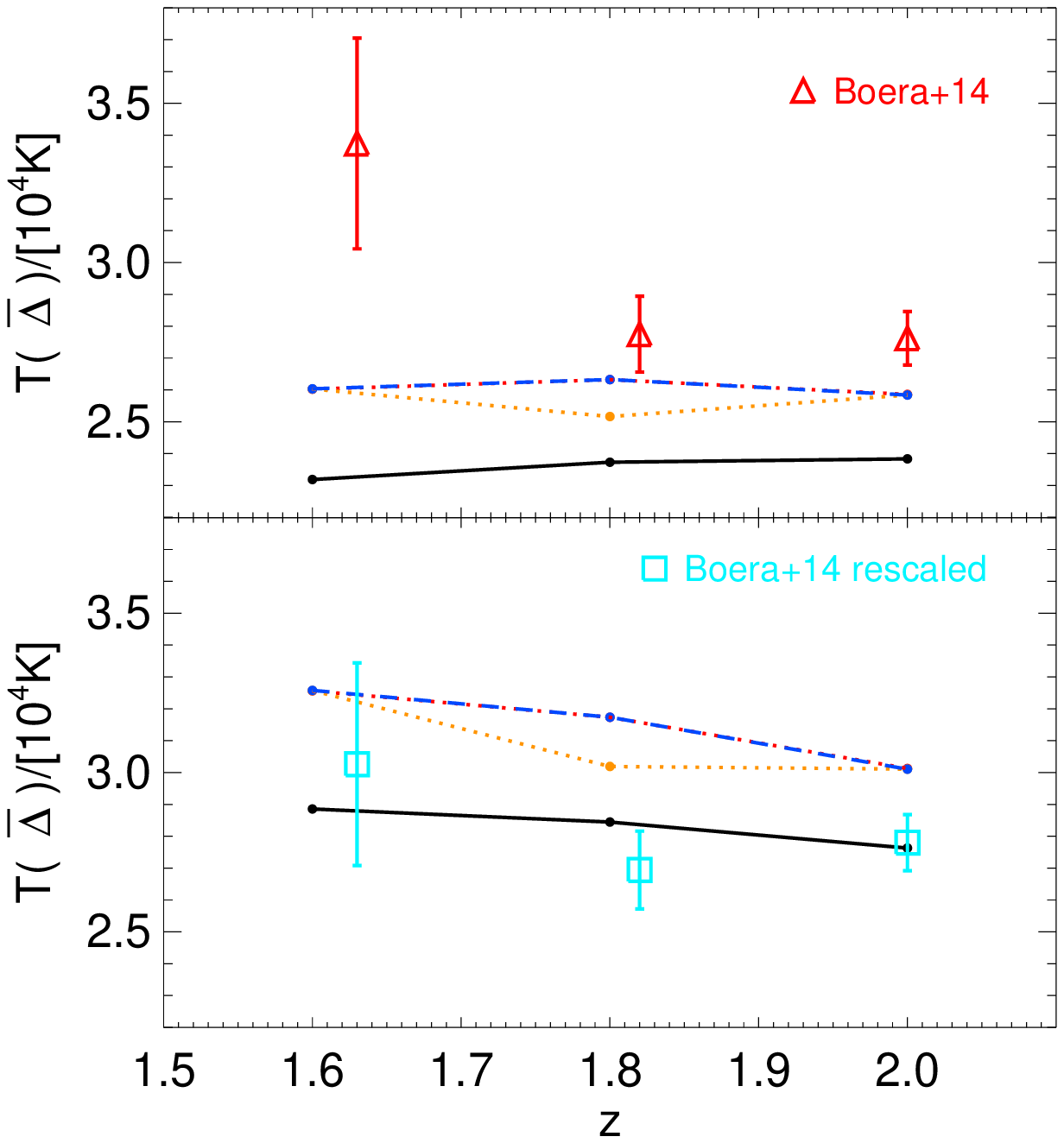}        
\end{minipage}
\vspace{-0.0cm}
\caption{{ \it Left:} The redshift evolution of the volume weighted
  temperature at mean density, $T_0$ (upper panel) and slope of the
  power-law temperature-density relation, $\gamma-1$ (lower panel),
  where $T=T_{0}\Delta^{\gamma-1}$. The parameters are obtained from
  the volume weighted $T-\Delta$ plane by finding the mode of the gas
  temperature in bins of width $0.02$ dex at $\log\Delta=0$ and
  $-0.7$. The $T_0$ data points with 1$\sigma$ errors (red triangles)
  are inferred from IGM temperature measurements at the
    characteristic density probed by the \Lya forest,
    $T(\bar{\Delta})$, obtained by \citet{Boera_2014}, evaluated at
    the $\gamma-1$ values predicted by the 80-1024 model. The same
  measurements scaled to match the \citet{Becker_2011MNRAS} effective
  optical depth are displayed as cyan squares (see text for
  details). The $\gamma-1$ data points with $1\sigma$ errors are from
  \citet{Schaye_2000MNRAS} (orange inverted triangle) and
  \citet{Ricotti_2000ApJ} (green diamond).{ \it Right:} The
    temperature at the characteristic density probed by the \Lya
    forest, $T(\bar{\Delta})$, for the same models overlaid with data
    from \citet{Boera_2014} (top panel) and following recalibration to
    match the \citet{Becker_2011MNRAS} effective optical depth (bottom
    panel).  The characteristic densities at $z=[1.63,\,1.82,\,2.00]$
    are $\bar{\Delta}=[5.13,\,4.55,\,4.11]$
    ($\bar{\Delta}=[7.65,\,6.32,\,5.38]$) for the fiducial
    (recalibrated) \citet{Boera_2014} temperature measurements. }
\label{fig:ToGamma}
\end{figure*}

The evolution of the low-density IGM thermal history at $z<2$ in a
sub-set of the simulations is displayed in Figure~\ref{fig:ToGamma}.
This is parameterised as a power law temperature-density ($T-\Delta$)
relation for the low density ($\Delta<1$--$10$) IGM, $T=T_0
\Delta^{\gamma-1}$.  Here $\Delta=\rho/\langle\rho\rangle$ is the gas
density normalised by the background density, $T_0$ is the gas
temperature at mean density and $\gamma-1$ is the slope of the
relation \citep{Gnedin_1998MNRAS,Sanderbeck_2016MNRAS}.  Note that the
80-512-ps13, 80-512-ps13+agn and 80-512 models have very similar
thermal histories; this suggests gas at the mean background density is
not strongly impacted by feedback. Additionally, by comparing these
models to the 80-1024 simulation (solid black line), it is clear that
the lower resolution simulations are not converged, with volume
weighted temperatures at mean density $\sim 800\rm\,K$ hotter in the
80-512 runs (further details may be found in the Appendix).

There are currently no detailed observational constraints on the
temperature of the low density IGM at $z<1.6$, but there are a few
measurements from optical data at $1.6\leq z \leq 2$. In the upper
left panel of Figure~\ref{fig:ToGamma} the red triangles show the IGM
temperature at mean density from \citet{Boera_2014}.  This study uses
the curvature of the \Lya forest transmitted flux to measure the gas
temperature at the characteristic density, $T(\bar{\Delta})$, probed
by the \Lya forest absorption.  We derive values for $T_0$ from these
data assuming the $\gamma-1$ for the 80-1024 model shown in the lower
left panel, such that $T_0=T(\bar{\Delta})/\bar{\Delta}^{\gamma-1}$.
Additionally, the cyan squares display the \citet{Boera_2014}
measurements obtained assuming the effective optical depth,
  \taueff\ (where \taueff$ = -\ln \langle F \rangle$ and $\langle F
  \rangle$ is the mean transmission in the \Lya forest), evolution
  used in an earlier curvature analysis performed by
  \citet{Becker_2011MNRAS} ({E. Boera, private communication}).  The
  effective optical depth sets the characteristic gas density probed
  by the \Lya forest, and this comparison gives an indication of the
  likely systematic uncertainty in $\tau_{\rm eff}$ arising from line
  of sight variance.  For completeness, we also directly compare the
  simulations to the $T(\bar{\Delta})$ measurements in the right panel
  of Figure~\ref{fig:ToGamma}. The measurements bracket all the
simulations at $z>1.6$.  The models are also formally consistent with
the constraints on $\gamma-1$ from \citet{Schaye_2000MNRAS} and
\citet{Ricotti_2000ApJ} shown in the lower left panel of
Figure~\ref{fig:ToGamma}.  These were obtained from an analysis of the
lower envelope of the column density -- velocity width plane obtained
from Voigt profile fits to the \Lya forest, although the error bars on
these measurements are very large.  Note also the simulations do not
follow non-equilibrium ionisation effects, and therefore produce a
temperature-density relation slope that is steeper by
$\Delta(\gamma-1)\sim 0.1$ at $z=1$--$2$ compared to the recent study
by \citet{Puchwein_2015MNRAS}.

In all cases the simulated $T_{0}$ monotonically decreases toward
lower redshift as the IGM cools due to adiabatic expansion, reaching
$\sim 4000\rm\,K$ by $z=0$.  The agreement with measurements of the
IGM thermal state at $1.6<z<2$ suggests the photo-heating rates used
in the UVB model are broadly correct at these (and higher) redshifts.
Significantly higher temperatures at mean density than those predicted
by the models at $z<1.6$ would require an additional, unidentified
source of heating in the low density IGM.

\subsection{Mock \Lya forest spectra}
\begin{figure*}
    \centering
    \begin{minipage}{.495\textwidth}
        \centering
        \includegraphics[trim={0.0cm 0.5cm 1.0cm 2.0cm}, clip=true, width=\columnwidth]{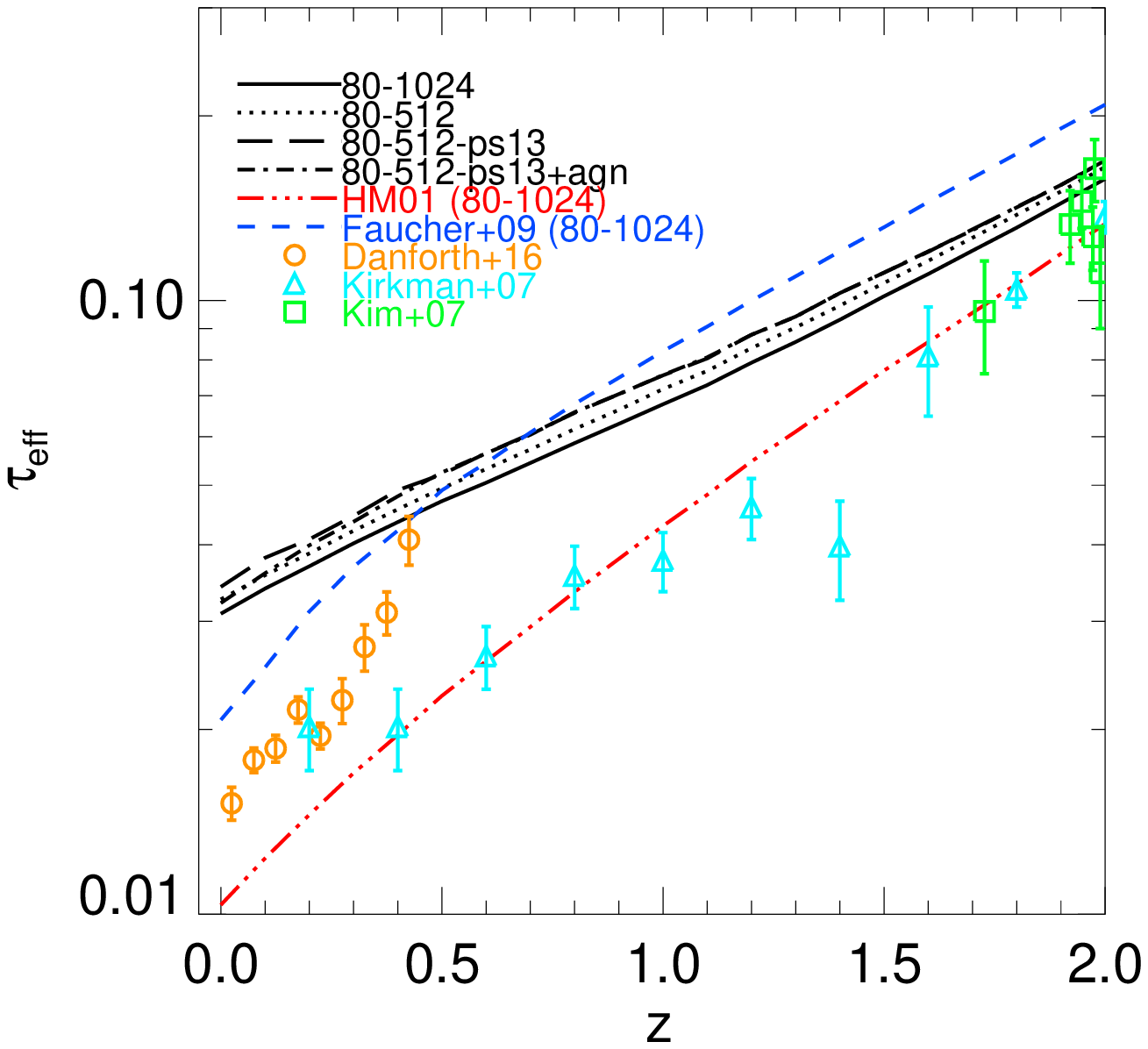}       
    \end{minipage}
    \begin{minipage}{.495\textwidth}
        \centering
        \includegraphics[trim={0.0cm 1.cm 1.6cm 2.2cm}, clip=true, width=\columnwidth]{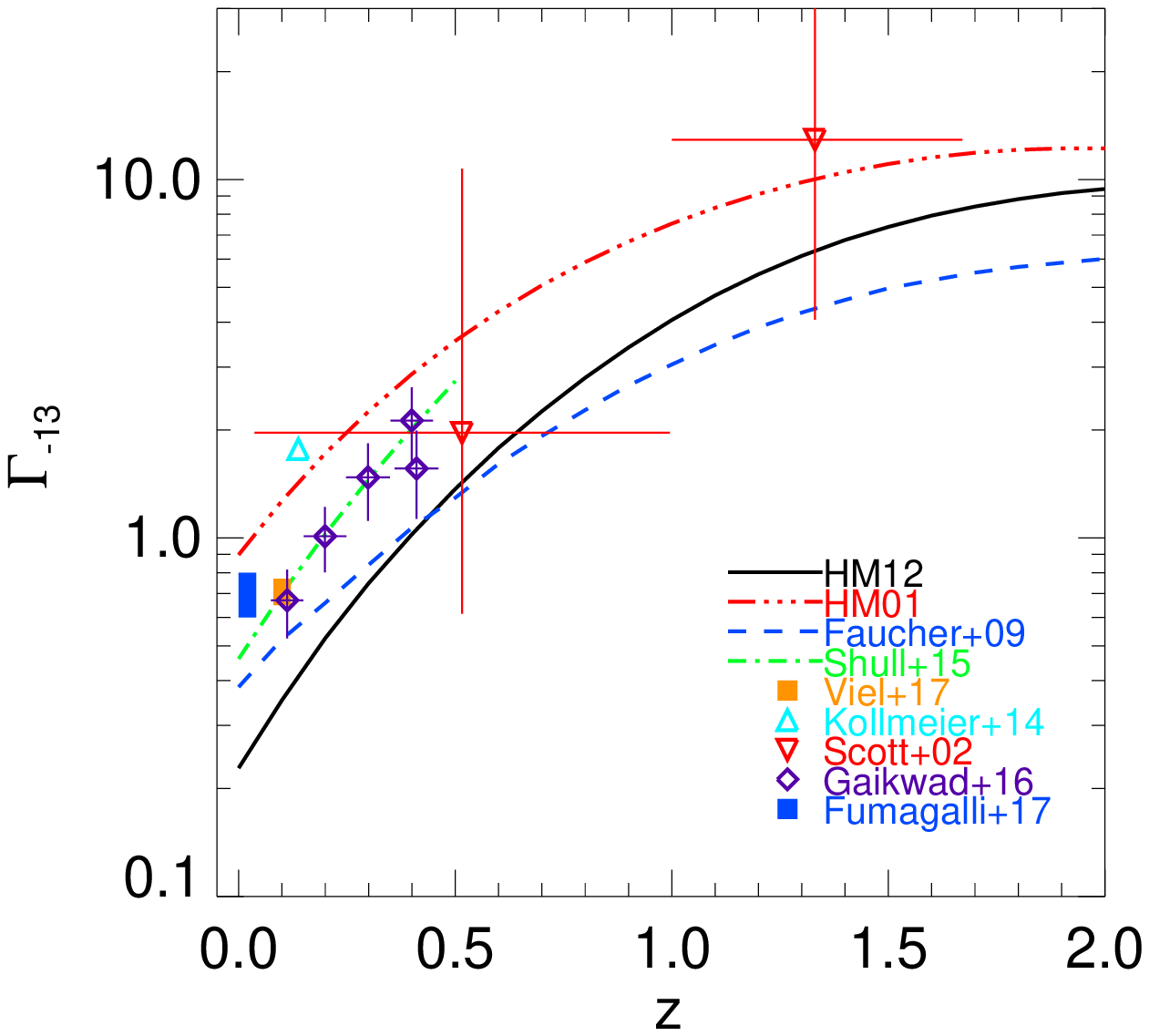}        
\end{minipage}
\vspace{-0.3cm}
\caption{{ \it Left:} The redshift evolution of the effective optical
  depth, $\tau_{\rm eff}$, predicted by the hydrodynamical
  simulations. The evolution using the HM01 (red triple-dot dashed
  line) and \citet{Faucher_2009ApJ} (blue dashed line) UVB models for
  emission from star forming galaxies and quasars are also shown for
  the rescaled 80-1024 model. The \Lya forest observations are from
  \citet{Kirkman_2007MNRAS}, \citet{Kim_2013A&A} and
  \citet{Danforth_2016ApJ}.  { \it Right:} The \HI photo-ionisation
  rate, $\Gamma_{\rm -13} = \Gamma_{\rm HI}/10^{\rm -13} {\rm s}^{\rm
    -1}$, evolution with redshift for the HM12 (black solid line),
  HM01 (red triple-dot dashed line) and \citet{Faucher_2009ApJ} (blue
  dashed line) UVB models. The observational measurements are from
  \citet{Scott_2002ApJ,Kollmeier_2014ApJ,Gaikwad_2016,Viel_2016} and
  \citet{Fumagalli_2017}.  The green dot-dashed line is the best-fit
  from \citet{Shull_2015ApJ}.}
\label{fig:obs_cmp}
\end{figure*}

Mock \Lya forest spectra were obtained by extracting $5000$ lines of
sight parallel to each of the box axes on regularly spaced grids of
$50^{2}$, $40^{2}$ and $30^{2}$, each with $2048$ pixels. The \HI \Lya
optical depths, $\tau^{\alpha}_{\rm HI}$, are obtained using the
method described by \citet{Theuns_1998MNRAS} combined with the Voigt
profile approximation described by
\citet{Tepper-Garcia_2006MNRAS}. The transmitted flux, $F$, in each
pixel is then $F=e^{- \tau^{\rm \alpha}_{\rm HI}}$, and the effective
optical depth evolution with redshift is averaged over all sight
lines.  In our analysis we also make use of the fact that the pixel
optical depths are inversely proportional to the \HI photo-ionisation
rate, $\tau^{\alpha}_{\rm HI} \propto \Gamma_{\rm HI}^{-1}$.  This
enables the optical depths (and hence the \taueff\ evolution) to be
rescaled to match different UVB models in post-processing
\citep[e.g.][]{Rauch_1997ApJ}.

In Figure~\ref{fig:obs_cmp} (left panel), the $\tau_{\rm eff}$
evolution for a variety of sub-grid physics and UVB models are
compared with \Lya forest observations.  The results shown by the
black curves use the HM12 UVB model, whereas the red triple-dot dashed and
blue dashed curves are obtained by scaling the simulated optical
depths to match the \HI photo-ionisation rates in the earlier
\citet[][HM01]{Haardt_2001} and \citet{Faucher_2009ApJ} UVB models
for emission from star-forming galaxies and quasars. The
evolution of \taueff\ in the 80-1024 model when using the HM01 UVB is
in reasonable agreement with observational measurements from
\citet{Kirkman_2007MNRAS} (except the dip at $z\simeq1.4$) and
\citet{Kim_2007MNRAS} at $1.7<z<2.0$, but underpredicts \taueff\ at
$z<0.4$.  In contrast, the fiducial HM12 UVB predicts \taueff\ values
that are too high, particularly $z<1.5$; a closely related
result was noted by \citet{Kollmeier_2014ApJ} when analysing the \HI
CDDF.  None of the UVB models is able to reproduce a
\taueff\ evolution that is in good agreement with the more recent COS
data from \citet{Danforth_2016ApJ} at $z\leq0.4$.  There is also
  a difference between these data and the measurements from
  \citet{Kirkman_2007MNRAS} at $z\sim 0.4$, possibly due to systematic
  differences in the measurements and cosmic variance. The
\citet{Faucher_2009ApJ} UVB is closer to these data than HM12 at
$z<0.5$, but overpredicts \taueff\ at all redshifts including $z=2$.
Interestingly, the differences between models with different feedback
implementations are small, indicating that it is the uncertain
amplitude of the UVB rather than galactic feedback that is largely
responsible for this discrepancy \citep[see also figure 1
  in][]{Dave_2010MNRAS}.

This is further illustrated in the right panel of
Figure~\ref{fig:obs_cmp} where observational estimates of the
\HI photoionisation rate, $\Gamma_{\rm -13} = \Gamma_{\rm HI}/10^{\rm
  -13} {\rm s}^{\rm -1}$, are compared to the three UVB models.  The
measurements by \citet{Shull_2015ApJ,Gaikwad_2016} and V17, based on
an analysis of the COS \Lya CDDF, are in mutual agreement and predict
photo-ionisation rates at $z<0.4$ intermediate between HM01 and HM12.
The value proposed by \citet{Kollmeier_2014ApJ} (cyan triangle) lies
very close to the HM01 UVB at $z=0.1$.  The red inverted
  triangles with large error bars display earlier constraints on
$\Gamma_{\rm HI}$ obtained from the proximity effect by
\citet{Scott_2002ApJ}.  The blue rectangle corresponds to the recent,
independent constraint from \citet{Fumagalli_2017} using observations
of H$\alpha$ fluorescence from \HI in a nearby galactic disk
\citep[see also][]{Adams2011}.

Differing assumptions regarding the relative contributions of
star forming galaxies and quasars and their spectral shape are
largely responsible for differences between existing UVB models. Given
these considerable differences, efforts to calibrate the \HI
photo-ionisation rate to match \taueff\ and CDDF measurements at
$z<0.4$ \citep[e.g.][]{Khaire2015,Madau2015}, as well as renewed
attempts to obtain measurements of $\Gamma_{\rm HI}$ in the redshift
interval $0.4< z < 2$, are desirable.


\section{Voigt profile analysis}
\label{sec:cddf_data}

We now proceed to present the results of the Voigt profile analysis of
our mock \Lya forest spectra.  We fit lines at three redshifts,
$z=[0.1,\,1,\,1.6]$. When comparing simulations to each other or
  to data compilations that have a range of resolution and
  signal-to-noise properties, the mock spectra are
post-processed by convolving them with an instrument profile
with a Full Width Half Maximum (FWHM) of $7$ \kms and rebinned to a
pixel size of 3 \kms.  A uniform Gaussian distributed noise is added
with a signal-to-noise of S/N$=50$ per pixel. The mean transmission of
the spectra are rescaled to match \taueff $=[0.012,\,0.043,\,0.086]$
at $z=[0.1,\,1,\,1.6]$. These values correspond to the 80-1024
  model \taueff\ evolution for the HM01 UVB for emission from
    star forming galaxies and quasars, which is in good agreement
  with the \citet{Kirkman_2007MNRAS} and \citet{Kim_2007MNRAS}
  $\tau_{\rm eff}$ measurements in the left panel of
  Figure~\ref{fig:obs_cmp}.

The exception to this is when we directly compare the simulations to
the V17 COS measurements of the CDDF and velocity width
distribution at $z=0.1$.  In this case, we follow V17 and convolve the
spectra with the LF1 G130M COS line spread function including
scattering at wavelength
$1350\,$\angstrom\footnote{\url{http://www.stsci.edu/hst/cos/performance/spectral_resolution/}}. The
spectra are rebinned to pixels of $\sim 7.2 \rm \, km\,s^{-1}$ and a
Gaussian distributed signal-to-noise of S/N$=30$ per resolution
element (or equivalently $\sim 20$ per pixel) is then added.  The
effective optical depth is scaled to \taueff$=0.021$ ($\langle F
\rangle =0.979$).  This corresponds to the value V17 required to match
the observed CDDF in the range $10^{13} \leq N_{\rm HI}/{\rm cm}^{-2}
\leq 10^{14}$.  

The COS observations of the CDDF and velocity width distribution at
$z=0.1$ are as described by V17; further details regarding the data
reduction and AGN spectra can also be found in \citet{Wakker_2015ApJ}
and Kim et al. (in prep).  The data comprises of 704 \HI lines within
column densities $10^{12.5} \leq N_{\rm HI}/\rm cm^{-2} \leq 10^{14.5}
$ with a total redshift path length of $\Delta z=4.991$, covering the
\Lya forest at $0<z<0.2$.  The Voigt profile fits are obtained using
\Lya absorption only.

The fitting of Voigt profiles to the mock \Lya spectra is performed
with \textsc{VPFIT} \citep{VPFIT}, which deconvolves the (already
  convolved) mock spectra with the instrument profile to obtain
  intrinsic line widths.  Importantly, this approach matches that
used to fit the COS observational data at $z=0.1$.  A total path
length of $10^5$\mpc\ is fitted for each model in segments of
$20$\mpc\ to obtain our simulated line lists. All lines within
$50\,$\kms\ of the start and end of each segment are ignored to avoid
spurious line fits due to edge effects.  Finally, to estimate the
sample variance in the COS observational data, we adopt the approach
of \citet{Rollinde_2013MNRAS} and bootstrap resample the mock spectra
with replacement over the observed path length $\Delta z=4.991$.  The
sampling is performed 1000 times over the total simulated path length
of $10^{5}h^{-1}\rm\,cMpc$.

\subsection{The CDDF}
\begin{figure}
\includegraphics[width=\columnwidth,trim={0.55cm 0.0cm 1.2cm 0.5cm},clip]{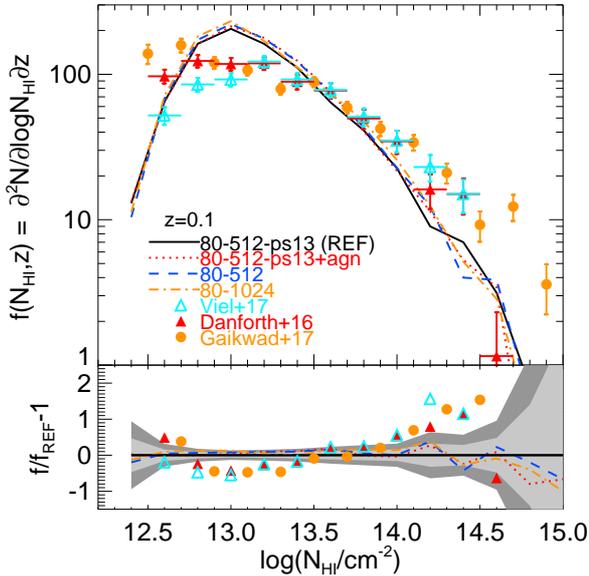}       
\vspace{-0.9cm}
\caption{The \HI column density distribution function at $z=0.1$ after
  scaling the mock spectra to a mean transmission $\langle F \rangle =
  0.979$. Three of the models use different prescriptions for
  following cold, dense gas and galactic feedback: \textsc{QUICKLYA}
  (blue dashed curve), multi-phase star formation and energy driven
  winds (black solid curve) and with the addition of AGN feedback (red
  dotted curve).  The fourth model (80-1024) uses \textsc{QUICKLYA}
  but has increased mass resolution.  The observational data points
  are from \citet{Danforth_2016ApJ} (red triangles),
    \citet{Gaikwad_2016} (orange circles) and V17 (cyan triangles).
  Only absorption lines with a relative error less than 50 per cent on
  the column density are used. The lower panel shows the simulations
  and COS observations relative to the reference model 80-512-ps13.
  The grey shaded areas display the sample variance estimate,
  corresponding to the 68 and 95 per cent confidence intervals when
  bootstrap resampling over a redshift path length of $\Delta
  z=4.991$.}
\label{fig:CDDF_COS} 
\end{figure}

\begin{figure*}
\includegraphics[width=.85\textwidth,trim={0.0cm 5.2cm 0.1cm 0.1cm},clip]{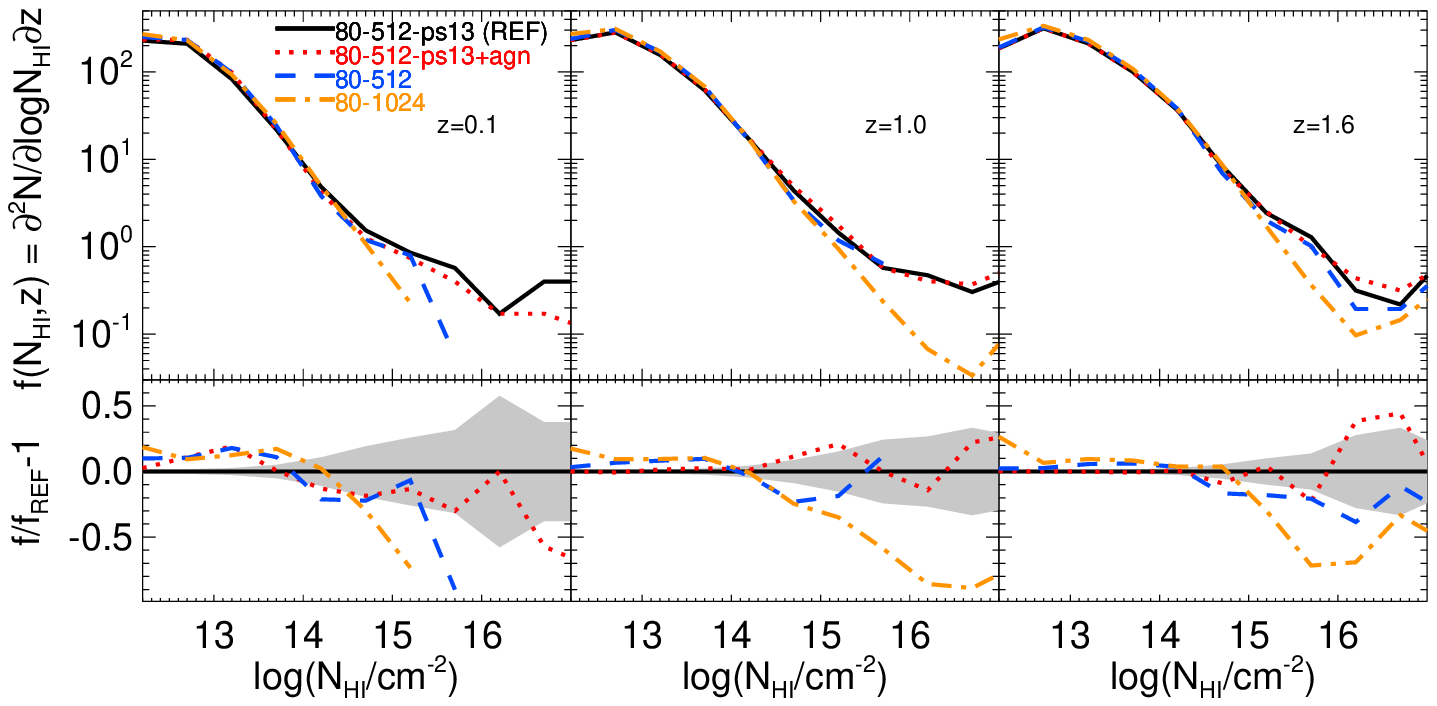}       
\vspace{-1.4cm}
\caption{{\it Upper panels:} The \Lya forest column density
  distribution function at $z=0.1,\,1,\,{\rm and}\,1.6$. The models
  shown here are the same as in Figure~\ref{fig:CDDF_COS}, except now
  the spectra are post-processed to have higher resolution ($\rm
  FWHM=7\rm\,km\,s^{-1}$) and signal-to-noise ($\rm S/N\sim 50$).
  Only lines with relative error less than 50 per cent on the column
  density are used. {\it Lower panels:} The CDDF relative to the
  reference model, 80-512-ps13. The grey shaded area shows the Poisson
  error for the reference model.}
\vspace{-0.5cm}
\label{fig:CDDF_Phys} 
\end{figure*}

Figure~\ref{fig:CDDF_COS} displays the comparison between the CDDF,
defined as the number of absorbers per unit log column density per
unit redshift, $f(N_{\rm HI},z)=\partial^2 N/\partial \log N_{\rm
  HI}\partial z$, from the Sherwood models and the COS observations
described by V17 at $z=0.1$ (see also figure 1 in V17).  Note that
only the 80-512-ps13+agn model was examined in detail by V17; in this
work we present a comparison of the additional Sherwood models in
which star formation, feedback and numerical resolution have been
changed while other parameters remain fixed.  We also display the
  independent measurements of the CDDF from \citet{Danforth_2016ApJ}
  (red triangles) and \citet{Gaikwad_2016} (orange circles), although
  we caution that differences in the line fitting methodologies used
  in these studies will bias any direct comparison with our
  simulations.

The simulations fail to capture the correct number of absorbers at
$N_{\rm HI}<10^{13.2}\rm\,cm^{-2}$, by as much as a factor $\sim2$ at
$10^{13}\rm\,cm^{-2}$.  Note that the simulations are well converged
with mass resolution -- this may be assessed by comparing the 80-512
(blue dashed curve) and 80-1024 (orange dot-dashed curve) models.
However, the CDDF is very sensitive to the signal-to-noise and
spectral resolution at low column densities
(cf. Figure~\ref{fig:CDDF_Phys} where S/N$=50$ and
FWHM=$7\rm\,km\,s^{-1}$), and the observational data are incomplete at
$N_{\rm HI}<10^{13}\rm\,cm^{-2}$. This discrepancy therefore reflects
differences between the true noise on the data and the idealised,
uniform noise added to the mock spectra and is probably not
significant.  In contrast, the agreement between the V17 data and
simulations at $10^{13.2}<N_{\rm HI}/\rm cm^{-2}<10^{14}$ is within
the $2\sigma$ uncertainty from sample variance, shown by the grey
shading in the lower panel.  At $N_{\rm HI}>10^{14}\rm\,cm^{-2}$,
however, the number of absorbers are underpredicted by a factor $\sim
2$ and the slope of the observed CDDF is shallower than the simulated
prediction. A similar discrepancy was present in the recent analysis
by \citet{Shull_2015ApJ} using simulations performed with the Eulerian
hydrodynamical code \textsc{enzo}, as well as in the study by
\citet{Gurvich_2016} with the moving mesh code \textsc{Arepo},
suggesting this discrepancy is not unique to \textsc{P-Gadget-3}
simulations. In contrast \citet{Gaikwad17} recently found better
  agreement with the CDDF at $N_{\rm HI}>10^{14}\rm\,cm^{-2}$ using
  \textsc{Gadget-2} adiabatic hydrodynamical simulations
  post-processed to incorporate radiative cooling, photo-heating and
  ionisation.  Note, however, the effect of gas pressure on the \HI
  distribution will not be correctly captured in these models,
  complicating interpretation of this result.

Lastly, it is clear that the various feedback scenarios explored by
the simulations impact very little overall on the CDDF at low column
densities, particularly at $N_{\rm HI}<10^{14}\rm\,cm^{-2}$.  Outflows
have a similarly small impact on the low column \Lya forest at $z>2$;
these systems are associated with the diffuse IGM rather than
circumgalactic gas \citep[][and see
  Section~\ref{sec:gas}]{Theuns2002,Viel_2013MNRAS}. Any differences
between the feedback models are within the expected 1$\sigma$ sample
variance.

This is further emphasised in Figure~\ref{fig:CDDF_Phys} where we
examine the \Lya absorption systems at column densities $
10^{12.2}\leq N_{\rm HI}/\rm{cm}^{-2} \leq 10^{17}$ at
$z=[0.1,\,1.0,\,1.6]$. Note that we adopt S/N$=50$ per pixel and a
instrument resolution $\rm FWHM=7$ \kms\ here. The lower panels
display the residual with respect to the 80-512-ps13 model, and grey
shading shows the Poisson error.  The \textsc{QUICKLYA} method
(80-512, blue dashed-line) agrees well with the more sophisticated
sub-grid physics models for weak absorption systems with $N_{\rm
  HI}<10^{13.5}\rm\,cm^{-2}$ at $z=0.1$ and $N_{\rm
  HI}<10^{14.5}\rm\,cm^{-2}$ at $z=1.6$.  However, higher column
density lines are either underpredicted or absent relative to the
other models.  Multi-phase star formation and stellar feedback
(80-512-ps13, black curve) results in a factor of 2-5 more absorption
systems at $N_{\rm HI}\ga 10^{14}\rm\,cm^{-2}$.  A similar result was
noted by \citet{Bolton_2017MNRAS} for the CDDF at $2<z<3$, although
the \HI column density threshold where \textsc{QUICKLYA}
underpredicted the CDDF was higher, $N_{\rm HI}\simeq
10^{14.5-15}\rm\,cm^{-2}$.  The lack of $N_{\rm HI}\ga
10^{14}\rm\,cm^{-2}$ absorbers is in part because of missing cold
dense gas in the \textsc{QUICKLYA} model, but also because winds
redistribute gas into the diffuse IGM and introduce peculiar
velocities that can increase the equivalent widths of the lines
\citep[][]{Dave_2010MNRAS,Meiksin2015}.

Finally, including AGN feedback in the Sherwood models yields similar
results for the CDDF relative to the stellar feedback only model,
except at $z=0.1$ where the incidence of high column density systems
is slightly lower (although the total number of absorbers here is
small).  A stronger effect was noted by \citet{Gurvich_2016}, who
found that the more vigorous AGN feedback in the Illustris simulation
suppresses the CDDF at $N_{\rm HI}=10^{14}$--$10^{15}\rm\,cm^{-2}$ by
a factor of two due to changes in the density and temperature of the
gas associated with these absorbers.  The higher temperatures reduce
the \Lya opacity by lowering the radiative recombination rate and (for
$T\ga 10^{5}\rm\,K$) collisionally ionising the gas. In contrast,
\citet[][see their figure C1]{Tepper-Garcia_2012MNRAS} find stellar
and AGN feedback have a negligible effect on the CDDF at $z=0.25$
using simulations drawn from the OWLS project \citep{Schaye2010}.
These differences likely reflect variations in the sub-grid physics,
as the precise impact of feedback will depend on how efficiently gas
is ejected from dark matter haloes.

In contrast to the present study, \citet{Tepper-Garcia_2012MNRAS} also
found reasonably good agreement with the \HI CDDF measurement at
$10^{14}\leq N_{\rm HI}/\rm cm^{-2} \leq 10^{14.5}$ from
\citet{Lehner_2007ApJ} based on an earlier compilation of FUSE
(\emph{Far Ultraviolet Spectroscopic Explorer}, FWHM$\sim20
\rm\,km\,s^{-1}$) and HST/STIS (\emph{Space Telescope Imaging
  Spectrograph}, FWHM$\sim 7\rm\,km\,s^{-1}$) observations at $z \la
0.4$. However, the models in \citet{Tepper-Garcia_2012MNRAS}
underpredicted the incidence of saturated lines at $N_{\rm HI}\sim
10^{14.5}\rm\,cm^{-2}$; their \Lya only line fits underestimated the
column density of these absorbers since the \citet{Lehner_2007ApJ}
HST/STIS data are obtained from \Lya and higher order Lyman lines.
The \citet{Lehner_2007ApJ} \HI CDDF furthermore only includes
absorbers with velocity widths $b_{\rm HI}<40$\kms\ and relative
errors less than 40 per cent on $N_{\rm HI}$ and $b_{\rm HI}$
\citep[i.e. figure 14 in][]{Lehner2008erratum}.  We have confirmed we
obtain similar agreement with the \citet{Lehner_2007ApJ} CDDF
measurements and our simulations (for $\rm S/N=50$ and FWHM=$7\rm\,
km\,s^{-1}$) to that obtained by \citet{Tepper-Garcia_2012MNRAS} when
selecting only absorbers with $b_{\rm HI}<40\rm\,km\,s^{-1}$.

\subsection{The velocity width distribution}

\begin{figure}
\includegraphics[width=\columnwidth,trim={0.4cm 0.0cm 1.2cm 0.5cm},clip]{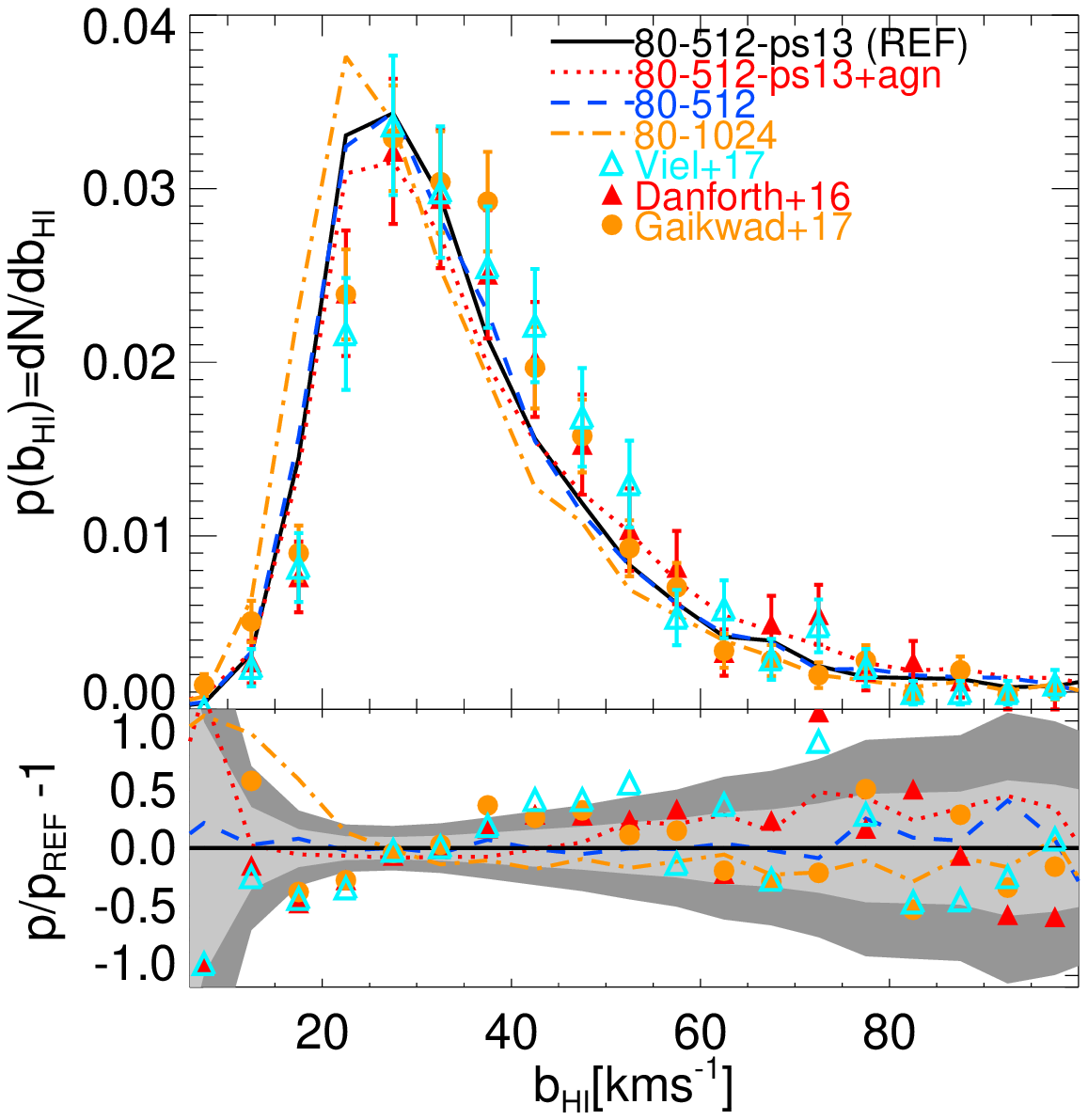}       
\vspace{-0.9cm}
\caption{The \Lya velocity width distribution at $z=0.1$ after scaling
  the mock spectra to a mean transmission $\langle F \rangle =
  0.979$. The observational data points are from
  \citet{Danforth_2016ApJ} (red triangles), \citet{Gaikwad_2016}
    (orange circles) and V17 (cyan triangles).  Only lines with
  $N_{\rm HI}=10^{13}$--$10^{14}\rm\,cm^{-2}$ and a relative error
  less than 50 per cent on the velocity widths are included. The lower
  panel shows the simulations and COS observations relative to the
  reference model 80-512-ps13.  The grey shaded areas display the
  sample variance estimate, corresponding to the 68 and 95 per cent
  confidence intervals when bootstrap resampling over a redshift path
  length of $\Delta z=4.991$.}
\label{fig:BDist_COS} 
\end{figure}

\begin{figure*}
\includegraphics[width=.85\textwidth,trim={0.cm 6.3cm 0.2cm 0.0cm},clip]{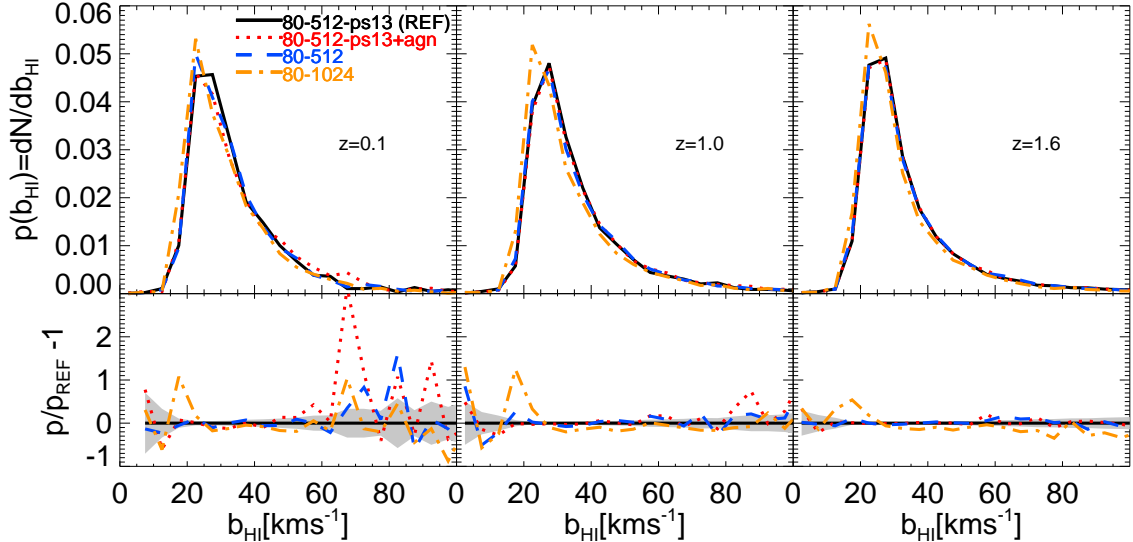}       
\vspace{-0.3cm}
\caption{The \Lya forest velocity width distribution at
  $z=0.1,\,1,\,{\rm and}\,1.6$. The curves are as for
  Figure~\ref{fig:CDDF_Phys}.  {\it Lower panels:} The velocity width
  distribution relative to the reference model, 80-512-ps13. Only
  lines with $N_{\rm HI}=10^{13}$--$10^{14}\rm\,cm^{-2}$ and a
  relative error less then 50 per cent on velocity widths are
  used. The grey shaded area shows the Poisson error for the reference
  model.}
\vspace{-0.5cm}
\label{fig:BDist_Phys} 
\end{figure*}

\begin{figure}
\includegraphics[width=\columnwidth,trim={1.2cm 0.2cm 0cm 0cm},clip]{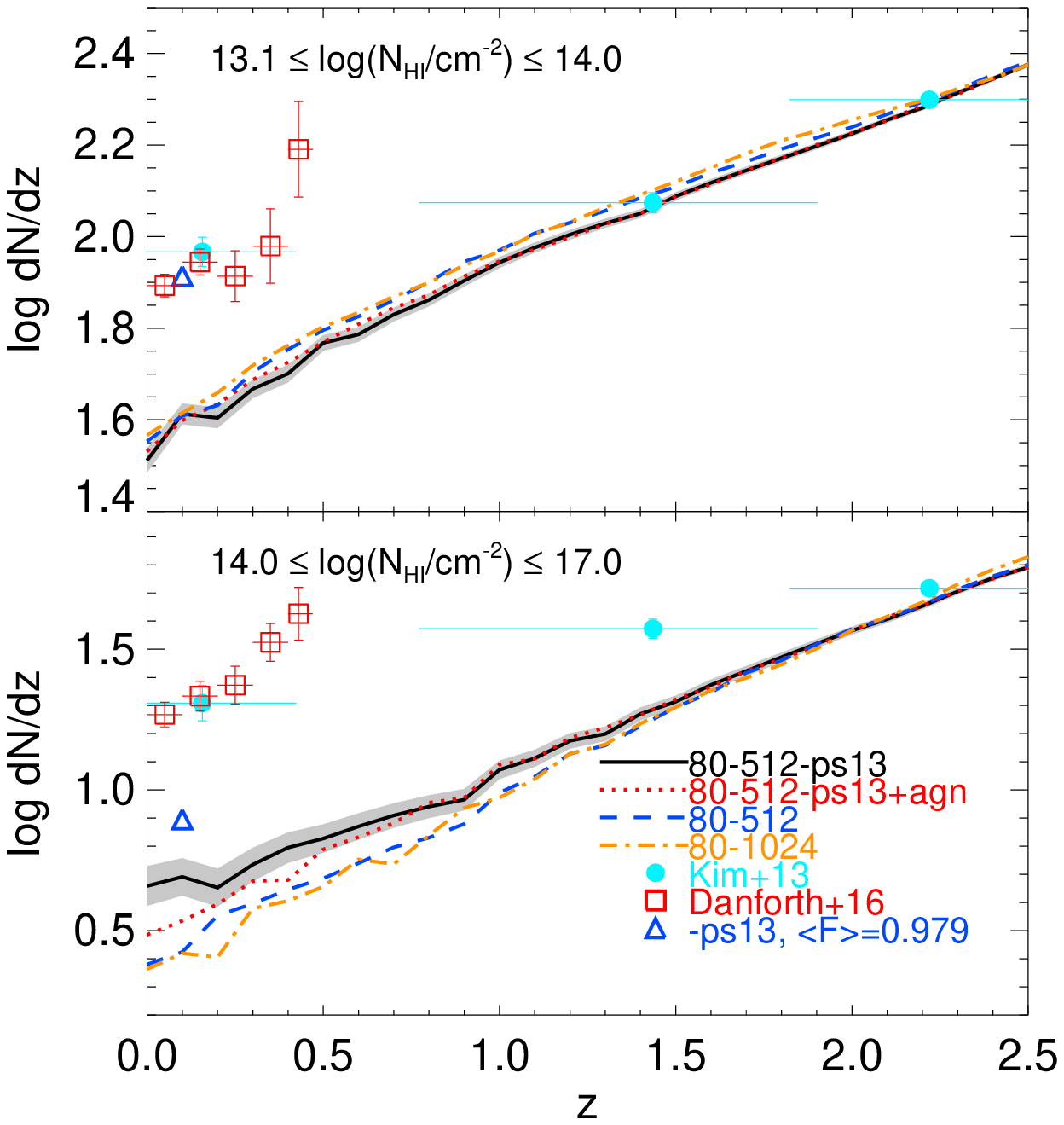}
\vspace{-0.5cm}
\caption{Redshift evolution in the number of \Lya absorbers per unit
  redshift for a subset of the simulations. Absorbers with column
  densities $10^{13.1}\leq N_{\rm HI}/{\rm cm}^{-2} < 10^{14}$
  (top-panel) and $10^{14} \leq N_{\rm HI}/{\rm cm}^{-2} \le 10^{17}$
  (bottom-panel) are shown. The observational data points with Poisson
  error bars are from \citet{Danforth_2016ApJ} (open red squares) and
  a data compilation presented by \citet{Kim_2013A&A} (filled cyan
  circles, see text for details). The blue triangles show the
  80-512-ps13 model rescaled to $\langle F \rangle=0.979$ at
  $z=0.1$. The grey shaded area shows the Poisson errors for the
  80-512-ps13 model.}
\vspace{-0.5cm}
\label{fig:dndz}
\end{figure}

We next turn to consider the distribution of \Lya line velocity
widths, $b_{\rm HI}$.  Figure~\ref{fig:BDist_COS} displays a
comparison between the V17 COS measurements and simulations at
$z=0.1$.  We again show the independent measurements from
  \citet{Danforth_2016ApJ} (red triangles) and \citet{Gaikwad_2016}
  (orange circles) for comparison.  Overall, the models slightly
underpredict the incidence of broad lines at $b_{\rm HI}=40$--$60$
\kms\ and overpredict at $b_{\rm HI}<30$\kms, although generally the
data lie within the expected 95 per cent scatter arising from sample
variance in the reference model.  It is important to note, however,
that unlike the CDDF (where the effect of feedback is comparable to or
smaller than the effect of mass resolution) the line widths are not
well converged with mass resolution (see also the Appendix later). The
distribution from the 80-1024 simulation (orange dot-dashed curve)
differs significantly from the 80-512 model (blue dashed curve),
particularly at $b_{\rm HI}\la 30\,$\kms, with a median value of
$30.8$\kms\ (cf. $33.5$\kms\ for the 80-512 model and $36.2$\kms\ for
the COS data).  Higher resolution simulations will therefore predict
line widths that are narrower than observed.  As pointed out by V17,
this suggests there may be additional physics that broadens the
absorption lines that is not incorporated into the simulations at
present.

We may nevertheless still examine the relative difference between the
different feedback models.  The incorporation of multi-phase star
formation and supernovae-driven winds (black curve) makes very little
difference compared to the \textsc{QUICKLYA} model.  The addition of
AGN feedback (red dotted curve) also has a small effect, increasing
the number of broad lines at $b_{\rm HI}>50\rm\,km\,s^{-1}$ with a
median\footnote{This is consistent with V17, who reported an increase
  by roughly $2$ \kms\ in the median b-parameter when including AGN
  feedback.  Note that V17 incorrectly reported this quantity as the
  peak of the velocity width distribution rather than the median.}
b-parameter $35.3$\kms.

In comparison to earlier work, \citet{Dave_2010MNRAS} also reported a
small difference of $\sim1$\kms\ in the median $b_{\rm HI}$ values
predicted by their constant wind and momentum-driven wind models at
$z=0$. However, their velocity widths are generally higher then we
find here, with a median value of $\sim43$\kms.  The mass resolution
of their simulations is around factor of two higher than this work
($M_{\rm gas}=2.49\times 10^{7}h^{-1}M_{\odot}$) although performed in
smaller volumes $\,(48h^{-1}\rm\,cMpc)^{3}$. This difference is
therefore most likely because \citet{Dave_2010MNRAS} do not
  deconvolve their Voigt profile fits with the instrument profile (see
  the first paragraph in section 4 of that paper).
\citet{Tepper-Garcia_2012MNRAS} similarly find that AGN feedback
produces a small increase in the typical \Lya line widths at $z=0.25$
using the OWLS simulations, and report a median b-parameter of
$29.4\,$\kms\ from mock spectra with FWHM=$7$\kms\ and a S/N$=50$ per
pixel. This is similar to the value of $27.7$\kms\ we obtain from our
AGN feedback (80-512-ps13+agn) model using the same signal-to-noise
and instrument resolution.  As \citet{Tepper-Garcia_2012MNRAS} also
use SPH simulations performed at slightly lower mass resolution
compared to this work ($M_{\rm gas}=8.7\times 10^{7}h^{-1}M{\odot}$),
their line widths will be similarly under-resolved.  More recently,
\citet{Gaikwad17} also found their simulated velocity width
distribution predicts lines that are narrower than the COS data.  It
is possible that the lack of any coupling between radiative cooling,
photo-ionisation and the hydrodynamical response of the gas in their
simulations may exaggerate this difference further, however.

The impact of feedback on the velocity distribution at
$z=[0.1,\,1.0,\,1.6]$ is displayed in Figure~\ref{fig:BDist_Phys}, for
FWHM$=7\rm\,km\,s^{-1}$ and $\rm S/N=50$ per pixel. Again it is clear
that the velocity width distribution is not converged with mass
resolution.  By comparing the 80-1024 (orange dot dashed line) and
80-512 (blue dashed line) the line widths differ significantly at
$b_{\rm HI}<30\rm\,km\,s^{-1}$. This implies a comparison between
simulations and observations requires a mass resolution of at least
$m_{\rm gas} \simeq 6\times 10^{6}h^{-1}M_{\odot}$. However, as for
the CDDF, the simulations with varying sub-grid physics performed at
the same mass resolution are very similar.  The only exception is at
$z=0.1$ when hot gas from AGN feedback produces slightly broader
lines.

\subsection{The evolution of number density of absorbers} \label{sec:dndz}

Lastly we examine the number of \HI absorbers per unit redshift,
$dN/dz$.  Figure~\ref{fig:dndz} displays $dN/dz$ for low ($10^{13.1}
\le N_{\rm HI}/\rm{cm}^{-2} \leq 10^{14}$, upper panel) and high
($10^{14} < N_{\rm HI}/\rm{cm}^{-2}\leq 10^{17}$, lower panel) column
density absorbers, overlaid with a compilation\footnote{The individual
  measurements are from
  \citet{Hu1995,Lu1996,Kim1997,Kirkman1997,Savaglio1999,Kim2001,Sembach2004,Williger2006,Aracil2006,Janknecht_2006A&A,Lehner_2007ApJ,Williger_2010MNRAS}.}
of observational data from \citet{Kim_2013A&A}. In this instance we do
not attempt to match the resolution and signal-to-noise of the
simulations to the data compilation, which was obtained with a
  variety of instruments.  We instead use FWHM$=7\,$\kms\ and
S/N$=50$ per pixel and assume the HM01 UVB for emission from star
  forming galaxies and quasars.  We also show the more recent
  COS measurements\footnote{Note the \citet{Danforth_2016ApJ}
      measurements in the upper panel of Fig.~\ref{fig:dndz} are for
      the slightly different column density range $10^{13} \le N_{\rm
        HI}/\rm{cm}^{-2} \leq 10^{14}$, rather than $10^{13.1} \le
      N_{\rm HI}/\rm{cm}^{-2} \leq 10^{14}$ used for the
      \citet{Kim_2013A&A} data compilation.}  from
  \citet{Danforth_2016ApJ} at $0\leq z \leq0.47$.  These are formally
  consistent with the earlier data compilation at $z\leq 0.3$, but
  with an increased incidence of lines at $0.3< z<0.5$. Note also we
use \Lya only fits for comparison to the observations, which may
complicate the comparison to the stronger, saturated lines with column
densities inferred using higher order Lyman lines.  We have therefore
tested this by also directly integrating the \HI number densities
along the simulated sight-lines to obtain $dN/dz$ \citep[see
  also][]{Gurvich_2016}.  This procedure yields similar results to the
Voigt profile fits.

The number of low column density \Lya forest absorbers (top panel) is
not sensitive to the differences in star formation and feedback
prescriptions in the models.  This is consistent with the CDDF, where
galactic winds and AGN feedback do not significantly impact on the gas
distribution at the low column densities that probe the diffuse
  IGM \cite[see also][]{Dave_2010MNRAS}.  The simulations are in good
agreement with the observational data at $z>1$, but underpredict the
number of weak lines at $z=0.1$.  To investigate this further, we have
rescaled the mean transmission in the 80-512-ps13 model to $\langle F
\rangle=0.979$ at $z=0.1$ (blue triangle), which V17 find provides a
better match to CDDF at $10^{13} < N_{\rm HI}/\rm{cm}^{-2}
<10^{14}$. The improved agreement again demonstrates this discrepancy
is due to the uncertain UVB amplitude; the HM01 model overpredicts the
\HI photo-ionisation rate at $z\la 0.5$.

In contrast to the low column density systems, the simulated
predictions for higher column density \Lya forest absorbers exhibit
some differences with the choice of star formation and feedback model
at $z<1.5$. The 80-512-ps13 model produces a factor of $\sim2$
higher $dN/dz$ relative to 80-512 at $z\simeq0$, with the AGN feedback
intermediate between the two.  However, all models fail to reproduce
the observational data at $z\leq 1.5$. Rescaling the mean transmission
at $z=0.1$ slightly improves agreement but does not remove the
discrepancy.  In combination with the CDDF, this strongly suggests the
simulations are not correctly capturing the saturated \Lya absorption
systems that are still optically thin to Lyman continuum radiation.


\section{Identifying the gas responsible for low redshift \Lya forest absorbers} \label{sec:gas}

\begin{table*}
\caption{A comparison of baryon phases in the simulations with
  different treatments for cold, dense gas and feedback at $z=0\,,0.1$
  and $1.6$ where $\Delta_{\rm th}=97.2,\,65.9$ and $62$,
  respectively. The percentage of the total gas mass is divided into
  four categories following \citet{Dave_2010MNRAS}. } \centering
\begin{tabular}{clcccccc}
\hline

Name             & Model & Diffuse & WHIM & Hot Halo & Condensed & Stars\\               
                 &       &$\Delta<\Delta_{\rm th},T<10^5$K&$\Delta<\Delta_{\rm th},T>10^5{\rm K}$&$\Delta >\Delta_{\rm th},T>10^5$K&$\Delta>\Delta_{\rm th},T<10^5$K\\
  \hline     
  $z=0.1$        & 80-512          &36.5    &21.3 &17.1 &0.9 &24.2\\         
                 & 80-512-ps13     &36.6    &24.3 &22.4 &9.1 &7.7\\         
                 & 80-512-ps13+agn &36.5    &36.9 &20.0 &3.3 &3.3\\  
       
\hline           
 $z=1.0$         & 80-512          &51.7   &15.4 &12.6 &1.9 &18.4\\       
                 & 80-512-ps13     &51.9   &17.0 &16.3 &10.6&4.3\\         
                 & 80-512-ps13+agn &51.6   &19.2 &17.7  &8.5&3.0\\ 
       
\hline
 $z=1.6$         & 80-512          &61.8   &11.3 &8.9  &2.8 &15.1\\         
                 & 80-512-ps13     &62.0   &12.3 &11.3 &11.5&2.8\\         
                 & 80-512-ps13+agn &62.0   &12.7 &11.7 &11.1&2.4\\      
       
\hline
\end{tabular}
\label{tab:feedback}
\end{table*}

\begin{figure*}
\includegraphics[width=0.85\textwidth,trim={0.2cm 8.7cm 0.0cm 1.2cm},clip]{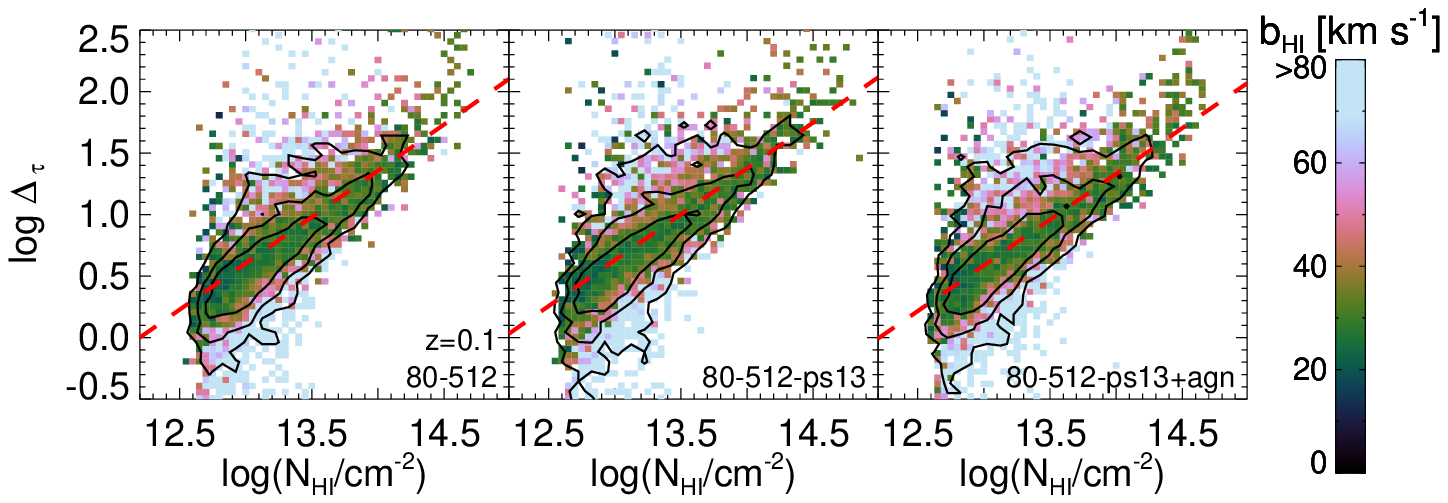} 
\includegraphics[width=0.85\textwidth,trim={0.2cm 8.7cm 0.0cm 1.2cm},clip]{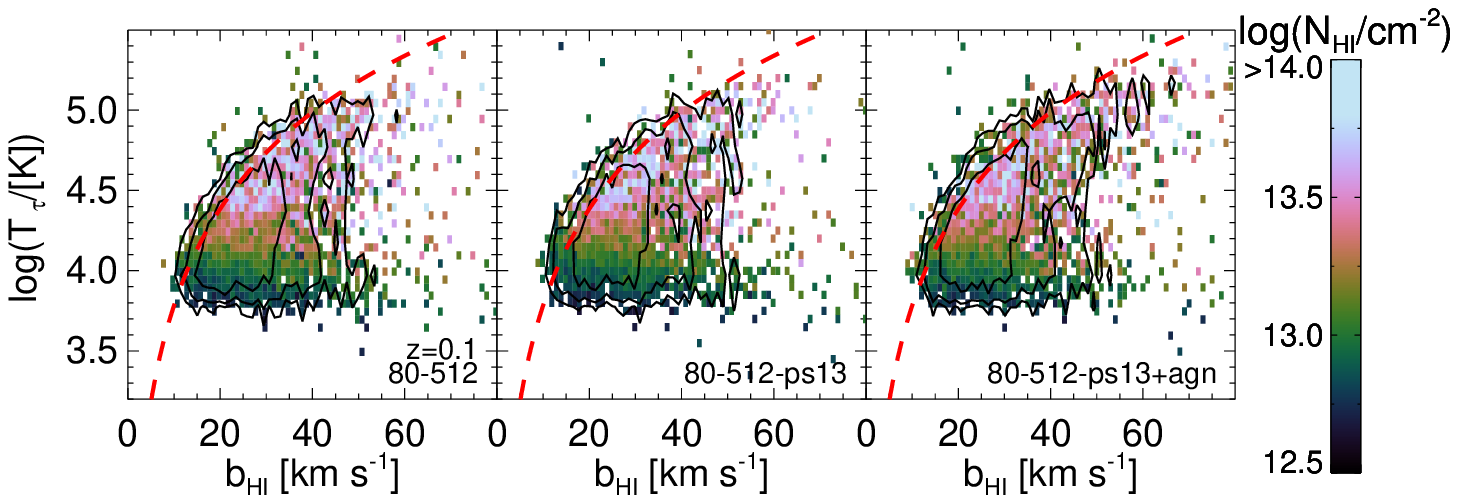}  
\vspace{-0.35cm}
\caption{{\it Upper panel:} The optical depth weighted normalised gas
  density against the \HI column density for simulations with
  different prescriptions for feedback at $z=0.1$ (from left to right,
  \textsc{QUICKLYA}, stellar feedback and stellar$+$agn feedback).
  The colours display the median velocity widths of the absorbers in
  each cell, and the black contours enclose 50, 75 and 90 per cent of
  total number of absorbers.  Only data points with relative errors on
  the velocity widths and \HI column densities less then 50 per cent
  are shown. The red dashed line shows the analytical model of
  \citet{Schaye_2001ApJ}, evaluated at the gas temperature and \HI
  photo-ionisation rate used in the simulation (see
  Eq.~\ref{eq:NH1Delta}). {\it Lower panel:} The optical depth
  weighted gas temperature against the \Lya velocity widths, $b_{\rm
    HI}$.  The colours now correspond to the median \HI column density
  in each cell and the red dashed line shows the relationship between
  $T$ and $b_{\rm HI}$ for pure thermal broadening, given by
  Eq.~(\ref{eq:bH1T}).}
\label{fig:nhi_den} 
\end{figure*}

In order to more closely examine the physical origin of the absorption
in our models and assess the effect of feedback, we consider the
typical temperature and density of gas in the \textsc{QUICKLYA} model
(80-512) and the \citet{Puchwein_2013MNRAS} energy driven winds model
with (80-512-ps13+agn) and without (80-512-ps13) AGN feedback. We have
divided the gas into four categories based on temperature, $T_{\rm
  th}=10^{5}\rm\,K$, and normalised density thresholds, $\Delta_{\rm
  th}$, following \citet{Dave_2010MNRAS}.  These are i) diffuse gas
($\Delta<\Delta_{\rm th}, T<10^5$K) ii) the warm-hot IGM (WHIM,
$\Delta<\Delta_{\rm th}, T>10^5$K) iii) hot halo gas
($\Delta>\Delta_{\rm th}, T>10^5$K, and iv) condensed gas
($\Delta>\Delta_{\rm th}, T<10^5$K). In Table~\ref{tab:feedback}, the
percentage of the total baryon mass in each category is shown at
$z=0.1,\,1$ and $1.6$ where $\Delta_{\rm th}=97.2,\,65.9$ and
  $62$, respectively \citep[here $\Delta_{\rm th}$ is computed using
  equations 1 and 2 in][]{Dave_2010MNRAS}.  We have also included the
percentage of gas that has been converted into collsionless star
particles.

The redshift evolution of gas in the diffuse phase is very similar for
all models, with no noticeable difference due to feedback. This
diffuse gas follows a power law temperature-density relationship,
$T=T_{0}\Delta^{\gamma-1}$ (see Figure~\ref{fig:ToGamma}).  However,
there is a significant increase in the mass fraction in the WHIM for
the model with AGN feedback at $z=0.1$, with a corresponding decrease
in the condensed phase fraction and stellar mass.  Including
multi-phase star formation and stellar feedback produces more gas in
the hot halo and condensed phases compared to the \textsc{QUICKLYA}
model at $z=0.1$.  In contrast, the \textsc{QUICKLYA} model produces
significantly more stars at $z\leq 1$ and -- while capturing the
diffuse IGM reasonably well -- will produce unreliable results for
gravitationally bound gas at $\Delta>\Delta_{\rm th}$. 

The mass fractions for the 80-512-ps13 model are broadly similar to
the ``vzw'' momentum driven winds model described in
\citet{Dave_2010MNRAS}, with slightly more ($\sim 5$ per cent) gas in
the dense phases with $\Delta>\Delta_{\rm th}$. There is also
reasonable agreement between the 80-512-ps13 model and the
post-processed adiabatic simulations described by \citet{Gaikwad_2016}
at $z=0$, with around 4 per cent more (less) gas in the WHIM (hot
halo) phase.  This suggests the bulk of the hot gas with
$T>10^{5}\rm\,K$ in the models without AGN feedback is produced by
gravitational infall. The largest difference is the condensed phase
which contains 18.8 per cent of the mass at $z=0$ in
\citet{Gaikwad_2016}; this is because dense gas is not converted into
stars in their adiabatic simulations.

The connection between the density and temperature of the gas and the
column density and velocity widths of the \Lya absorbers at $z=0.1$ is
summarised in Figure~\ref{fig:nhi_den}.  Along each simulated line of
sight we have computed the optical depth weighted temperature, $T_{
  \tau}$, and density, $\Delta_{ \tau}$, where $X_{\tau,\rm
  j}=\sum_{\rm i}\tau_{\rm i,j}X_{\rm i}/\sum_{\rm i} \tau_{\rm i,j}$
and $\tau_{\rm i,j}$ is optical depth at the $i^{\rm th}$ pixel in
real space contributing to the $j^{\rm th}$ pixel in velocity space
\citep{Schaye_1999MNRAS}.  We then associate $T_{ \tau}$ and $\Delta_{
  \tau}$ with \nhi\ and \bhi\ at the location of the line centres
identified by \textsc{VPFIT}.  This approach takes redshift space
distortions due to the peculiar motion of the gas into account.  As
before, following V17 we only consider systems with relative errors
less then 50 per cent on $b_{\rm HI}$ and $N_{\rm HI}$.

The top panel of Figure~\ref{fig:nhi_den} displays the relationship
between $N_{\rm HI}$ and $\Delta_{\rm \tau}$ in the simulations.  The
data are coloured according to their velocity widths. The red dashed
line shows the power-law relationship between \nhi and $\Delta$
predicted by \citet{Schaye_2001ApJ}, assuming the typical size of an
\HI absorber is the local Jeans scale:

\begin{equation} N_{\rm HI} = \frac{2.44\times 10^{13}\rm\,cm^{-2} }{\Gamma_{-13}} \left(\frac{\Delta}{10}\right)^{3/2} \left(\frac{T_{4}}{2}\right)^{-0.26} \left(\frac{1+z}{1.1}\right)^{9/2}. \label{eq:NH1Delta} \end{equation}

\noindent 
Here $\Gamma_{-13}=\Gamma_{\rm HI}/10^{-13}\rm\,s^{-1}$ is the \HI
photo-ionisation rate (where $\Gamma_{-13} = 0.78$ at $z=0.1$ in
  the 80-512-ps13 simulation for $\langle F \rangle =0.979$) and
$T_{4}=T/10^{4}\rm\,K$ is the gas temperature.

This provides a good description of the scaling between
gas density and \HI column density \citep[see
  also][]{Dave_1999ApJ,Dave_2010MNRAS,Tepper-Garcia_2012MNRAS}.  The
majority of the absorption lines at $N_{\rm
  HI}=10^{13}$--$10^{14}\rm\,cm^{-2}$ correspond to gas densities
$\Delta=4$--$40$ in the simulations (V17), with typical velocity
widths of $b_{\rm HI}=20$--$40\rm\,km\,s^{-1}$.  In all models there
are a small number of broad lines, $b_{\rm HI}>40\rm\,km\,s^{-1}$,
with column densities $N_{\rm HI}=10^{12.5}$--$10^{13.5}\rm\,cm^{-2}$
that are associated with hot $T\sim 10^{5}\rm\,K$ gas in the WHIM with
$\Delta \simeq 1$--$100$ \citep[see also][]{Tepper-Garcia_2012MNRAS}.
The effect of feedback on the diffuse, low column density gas is
rather subtle; there are slightly more broad lines lying above the
$N_{\rm HI}-\Delta_{\rm \tau}$ correlation for the AGN feedback
model, consistent with additional heating.  Although there is
significantly more hot gas in the WHIM by mass at $z=0.1$ when
including AGN feedback, much of this is also collisionally ionised.
In all instances, however, the higher column density systems, $N_{\rm
  HI}>10^{14}\rm\,cm^{-2}$ correspond to gas densities at $\Delta
  \ga 20$.  The simulations fail to correctly capture the number
density of these systems, some of which are at densities associated
with circumgalactic gas.

The lower panel in Figure~\ref{fig:nhi_den} shows the corresponding
values of $T_{\rm \tau}$ at the line centres against the velocity
widths. The red dashed line displays the expected velocity width under
the assumption of pure thermal broadening:

\begin{equation} b_{\rm HI} = 18.2 \rm\,km\,s^{-1} \left(\frac{T_{4}}{2}\right)^{1/2}. \label{eq:bH1T} \end{equation}

\noindent
In general, although the minimum line widths trace the expectation for
thermal broadening, most of the absorbers are significantly
  broader, particularly for low column density lines originating from
  low density gas \citep[see also][for a recent discussion at $z\sim
    3$]{Garzilli_2015MNRAS}.  Many of the absorbers are therefore
broadened through the additional effects of gas pressure (i.e. Jeans
smoothing).  The absorbers with $N_{\rm
  HI}=10^{13}$--$10^{14}\rm\,cm^{-2}$ are associated with gas with a
broad range of temperatures, but most are in the range
$T=10^{4}$--$10^{4.5}\rm\,K$.  The velocity widths of these lines
remain too narrow with respect to the COS data.  Again, the effect of
stellar and AGN feedback on the temperature of the absorbers is small.


\section{Conclusions}

We have investigated the impact of different feedback prescriptions on
the properties of the low redshift \Lya forest, using a selection of
hydrodynamical simulations drawn from the Sherwood simulation
suite. The simulations adopt stellar feedback, AGN feedback and a
simplified scheme for modelling the low column density \Lya forest
that converts all gas with $\Delta>1000$ and $T<10^{5}\rm\,K$ into
collisionless star particles.  We have examined the CDDF, velocity
width distribution, the redshift evolution in the line number density,
$dN/dz$, and the relationship between the column density, velocity
widths, and the underlying gas density and temperature.

Our analysis confirms the finding by V17 that the velocity widths of
simulated \Lya absorption lines with gas densities $\Delta=4$--$40$
are too narrow when compared with COS data.  We also stress that this
difference will be exacerbated further for simulations performed at
higher mass resolution than considered here; the line widths are
under-resolved at our fiducial resolution of $M_{\rm gas}=5.1\times
10^{7}h^{-1}M_{\odot}$ (see also the Appendix).  We also find
that absorbers with column densities $N_{\rm HI}>10^{14}\rm\,cm^{-2}$,
which correspond to gas with $\Delta \ga 20$ in our simulations,
are underestimated, in agreement with earlier work
\citep{Shull_2015ApJ,Gurvich_2016}.

On the other hand, using simulations that vary the feedback
prescription while keeping other parameters fixed, we find that the
impact of galactic winds and AGN feedback is generally very
modest. AGN feedback only impacts on the \Lya absorbers at very low
redshift, $z=0.1$, by producing some broader lines and slightly
reducing the incidence of systems with $N_{\rm HI}>10^{14}\rm cm^{-2}$
with respect to a model with only stellar feedback. There is no
noticeable impact on the CDDF due to stellar or AGN feedback at
$N_{\rm HI}<10^{14}\rm\,cm^{-2}$.  Furthermore, we find a simplified
scheme that ignores feedback and star formation altogether adequately
capture the properties of absorption lines with $N_{\rm
  HI}<10^{14}\rm\,cm^{-2}$, although the incidence of stronger systems
is underestimated compared to models that include feedback. Similarly,
the \Lya line number density and its evolution with redshift, $dN/dz$,
is in good agreement with observational data for low column density
systems, $10^{13.1}\leq N_{\rm HI}/\rm cm^{-2} \leq 10^{14}$, once the
amplitude of the UVB is rescaled to match the CDDF.  In contrast, for
high column density systems, $10^{14}\leq N_{\rm HI}/\rm cm^{-2} \leq
10^{17}$, the simulations underpredict the incidence of absorbers at
$z=0.1$ and $z=1$ by a factor of $\simeq 2.2$ and $3$, respectively.

This demonstrates that stellar and AGN feedback -- as currently
implemented in our simulations -- have a limited impact on the \Lya
forest.  We conclude that resolving the discrepancy between COS data
and the simulations requires an increase in the temperature of gas
with $\Delta=4$--$40$, either through (i) additional \HeII
photo-heating at $z>2$, (ii) additional fine-tuned feedback that
ejects overdense gas into the IGM at just the right temperature for it
to still contribute significantly to the \Lya forest or alternatively
(iii) a larger, currently unresolved turbulent component to the line
widths \citep{Oppenheimer2009,Iapichino2013,Gaikwad17}.  Note,
however, the first possibility would require a UVB with a harder
spectrum at $z>2$, potentially producing gas temperatures that are too
high with respect to existing observational constraints from the \Lya
forest at $1.6\leq z \leq 3$
\citep{Becker_2011MNRAS,Boera_2014,Bolton_2014MNRAS}.  Our results
provide further motivation for studying the impact of state-of-the-art
feedback models on the properties of the low redshift \Lya forest.

\section*{Acknowledgments}

The hydrodynamical simulations used in this work were performed with
supercomputer time awarded by the Partnership for Advanced Computing
in Europe (PRACE) 8th Call. We acknowledge PRACE for awarding us
access to the Curie supercomputer, based in France at the Tre Grand
Centre de Calcul (TGCC). This work also made use of the DiRAC High
Performance Computing System (HPCS) and the COSMOS shared memory
service at the University of Cambridge. These are operated on behalf
of the STFC DiRAC HPC facility. This equipment is funded by BIS
National E-infrastructure capital grant ST/J005673/1 and STFC grants
ST/H008586/1, ST/K00333X/1. We thank Volker Springel for making
\textsc{P-GADGET-3} available. FN is supported by a Vice-Chancellor's
Scholarship for Research Excellence. JSB acknowledges the support of a
Royal Society University Research Fellowship. MV and TSK are supported
by the FP7 ERC grant ``cosmoIGM'' and the INFN/PD51 grant. TSK
  also acknowledges the NSF-AST118913. MGH and EP acknowledge support
from the FP7 ERC Grant Emergence-320596, and EP gratefully
acknowledges support by the Kavli Foundation. DS acknowledges support
by the STFC and the ERC starting grant 638707 ``Black holes and their
host galaxies: co-evolution across cosmic time''.

\bibliographystyle{mnras}
\bibliography{bibliography}

\begin{thebibliography}{}
\makeatletter
\relax
\def\mn@urlcharsother{\let\do\@makeother \do\$\do\&\do\#\do\^\do\_\do\%\do\~}
\def\mn@doi{\begingroup\mn@urlcharsother \@ifnextchar [ {\mn@doi@}
  {\mn@doi@[]}}
\def\mn@doi@[#1]#2{\def\@tempa{#1}\ifx\@tempa\@empty \href
  {http://dx.doi.org/#2} {doi:#2}\else \href {http://dx.doi.org/#2} {#1}\fi
  \endgroup}
\def\mn@eprint#1#2{\mn@eprint@#1:#2::\@nil}
\def\mn@eprint@arXiv#1{\href {http://arxiv.org/abs/#1} {{\tt arXiv:#1}}}
\def\mn@eprint@dblp#1{\href {http://dblp.uni-trier.de/rec/bibtex/#1.xml}
  {dblp:#1}}
\def\mn@eprint@#1:#2:#3:#4\@nil{\def\@tempa {#1}\def\@tempb {#2}\def\@tempc
  {#3}\ifx \@tempc \@empty \let \@tempc \@tempb \let \@tempb \@tempa \fi \ifx
  \@tempb \@empty \def\@tempb {arXiv}\fi \@ifundefined
  {mn@eprint@\@tempb}{\@tempb:\@tempc}{\expandafter \expandafter \csname
  mn@eprint@\@tempb\endcsname \expandafter{\@tempc}}}

\bibitem[\protect\citeauthoryear{{Adams}, {Uson}, {Hill}  \&
  {MacQueen}}{{Adams} et~al.}{2011}]{Adams2011}
{Adams} J.~J.,  {Uson} J.~M.,  {Hill} G.~J.,   {MacQueen} P.~J.,  2011, \mn@doi
  [\apj] {10.1088/0004-637X/728/2/107}, 728, 107

\bibitem[\protect\citeauthoryear{{Aracil}, {Tripp}, {Bowen}, {Prochaska},
  {Chen}  \& {Frye}}{{Aracil} et~al.}{2006}]{Aracil2006}
{Aracil} B.,  {Tripp} T.~M.,  {Bowen} D.~V.,  {Prochaska} J.~X.,  {Chen} H.-W.,
    {Frye} B.~L.,  2006, \mn@doi [\mnras] {10.1111/j.1365-2966.2005.09962.x},
  367, 139

\bibitem[\protect\citeauthoryear{{Bahcall} et~al.,}{{Bahcall}
  et~al.}{1993}]{Bahcall1993}
{Bahcall} J.~N.,  et~al., 1993, \mn@doi [\apjs] {10.1086/191797}, 87, 1

\bibitem[\protect\citeauthoryear{{Becker}, {Bolton}, {Haehnelt}  \&
  {Sargent}}{{Becker} et~al.}{2011}]{Becker_2011MNRAS}
{Becker} G.~D.,  {Bolton} J.~S.,  {Haehnelt} M.~G.,   {Sargent} W.~L.~W.,
  2011, \mn@doi [\mnras] {10.1111/j.1365-2966.2010.17507.x}, 410, 1096

\bibitem[\protect\citeauthoryear{{Boera}, {Murphy}, {Becker}  \&
  {Bolton}}{{Boera} et~al.}{2014}]{Boera_2014}
{Boera} E.,  {Murphy} M.~T.,  {Becker} G.~D.,   {Bolton} J.~S.,  2014, \mn@doi
  [\mnras] {10.1093/mnras/stu660}, 441, 1916

\bibitem[\protect\citeauthoryear{{Bolton}, {Becker}, {Haehnelt}  \&
  {Viel}}{{Bolton} et~al.}{2014}]{Bolton_2014MNRAS}
{Bolton} J.~S.,  {Becker} G.~D.,  {Haehnelt} M.~G.,   {Viel} M.,  2014, \mn@doi
  [\mnras] {10.1093/mnras/stt2374}, 438, 2499

\bibitem[\protect\citeauthoryear{{Bolton}, {Puchwein}, {Sijacki}, {Haehnelt},
  {Kim}, {Meiksin}, {Regan}  \& {Viel}}{{Bolton}
  et~al.}{2017}]{Bolton_2017MNRAS}
{Bolton} J.~S.,  {Puchwein} E.,  {Sijacki} D.,  {Haehnelt} M.~G.,  {Kim} T.-S.,
   {Meiksin} A.,  {Regan} J.~A.,   {Viel} M.,  2017, \mn@doi [\mnras]
  {10.1093/mnras/stw2397}, 464, 897

\bibitem[\protect\citeauthoryear{{Carswell} \& {Webb}}{{Carswell} \&
  {Webb}}{2014}]{VPFIT}
{Carswell} R.~F.,  {Webb} J.~K.,  2014, {VPFIT: Voigt profile fitting program},
  Astrophysics Source Code Library (\mn@eprint {ascl} {1408.015})

\bibitem[\protect\citeauthoryear{{Cen} \& {Ostriker}}{{Cen} \&
  {Ostriker}}{1999}]{Cen1999}
{Cen} R.,  {Ostriker} J.~P.,  1999, \mn@doi [\apj] {10.1086/306949}, 514, 1

\bibitem[\protect\citeauthoryear{{Danforth} et~al.,}{{Danforth}
  et~al.}{2016}]{Danforth_2016ApJ}
{Danforth} C.~W.,  et~al., 2016, \mn@doi [\apj] {10.3847/0004-637X/817/2/111},
  817, 111

\bibitem[\protect\citeauthoryear{{Dav{\'e}}, {Hernquist}, {Katz}  \&
  {Weinberg}}{{Dav{\'e}} et~al.}{1999}]{Dave_1999ApJ}
{Dav{\'e}} R.,  {Hernquist} L.,  {Katz} N.,   {Weinberg} D.~H.,  1999, \mn@doi
  [\apj] {10.1086/306722}, 511, 521

\bibitem[\protect\citeauthoryear{{Dav{\'e}}, {Oppenheimer}, {Katz}, {Kollmeier}
   \& {Weinberg}}{{Dav{\'e}} et~al.}{2010}]{Dave_2010MNRAS}
{Dav{\'e}} R.,  {Oppenheimer} B.~D.,  {Katz} N.,  {Kollmeier} J.~A.,
  {Weinberg} D.~H.,  2010, \mn@doi [\mnras] {10.1111/j.1365-2966.2010.17279.x},
  408, 2051

\bibitem[\protect\citeauthoryear{{Faucher-Gigu{\`e}re}, {Lidz}, {Zaldarriaga}
  \& {Hernquist}}{{Faucher-Gigu{\`e}re} et~al.}{2009}]{Faucher_2009ApJ}
{Faucher-Gigu{\`e}re} C.,  {Lidz} A.,  {Zaldarriaga} M.,   {Hernquist} L.,
  2009, \mn@doi [\apj] {10.1088/0004-637X/703/2/1416}, 703, 1416

\bibitem[\protect\citeauthoryear{{Fumagalli}, {Haardt}, {Theuns}, {Morris},
  {Cantalupo}, {Madau}  \& {Fossati}}{{Fumagalli}
  et~al.}{2017}]{Fumagalli_2017}
{Fumagalli} M.,  {Haardt} F.,  {Theuns} T.,  {Morris} S.~L.,  {Cantalupo} S.,
  {Madau} P.,   {Fossati} M.,  2017, \mn@doi [\mnras] {10.1093/mnras/stx398},
  \href {http://adsabs.harvard.edu/abs/2017MNRAS.467.4802F} {467, 4802}

\bibitem[\protect\citeauthoryear{{Gaikwad}, {Khaire}, {Choudhury}  \&
  {Srianand}}{{Gaikwad} et~al.}{2017a}]{Gaikwad_2016}
{Gaikwad} P.,  {Khaire} V.,  {Choudhury} T.~R.,   {Srianand} R.,  2017a,
  \mn@doi [\mnras] {10.1093/mnras/stw3086}, 466, 838

\bibitem[\protect\citeauthoryear{{Gaikwad}, {Srianand}, {Choudhury}  \&
  {Khaire}}{{Gaikwad} et~al.}{2017b}]{Gaikwad17}
{Gaikwad} P.,  {Srianand} R.,  {Choudhury} T.~R.,   {Khaire} V.,  2017b,
  \mn@doi [\mnras] {10.1093/mnras/stx248}, 467, 3172

\bibitem[\protect\citeauthoryear{{Garzilli}, {Theuns}  \& {Schaye}}{{Garzilli}
  et~al.}{2015}]{Garzilli_2015MNRAS}
{Garzilli} A.,  {Theuns} T.,   {Schaye} J.,  2015, \mn@doi [\mnras]
  {10.1093/mnras/stv394}, 450, 1465

\bibitem[\protect\citeauthoryear{{Gnedin} \& {Hui}}{{Gnedin} \&
  {Hui}}{1998}]{Gnedin_1998MNRAS}
{Gnedin} N.~Y.,  {Hui} L.,  1998, \mn@doi [\mnras]
  {10.1046/j.1365-8711.1998.01249.x}, 296, 44

\bibitem[\protect\citeauthoryear{{Green} et~al.,}{{Green}
  et~al.}{2012}]{Green_2012ApJ}
{Green} J.~C.,  et~al., 2012, \mn@doi [\apj] {10.1088/0004-637X/744/1/60}, 744,
  60

\bibitem[\protect\citeauthoryear{{Gurvich}, {Burkhart}  \& {Bird}}{{Gurvich}
  et~al.}{2017}]{Gurvich_2016}
{Gurvich} A.,  {Burkhart} B.,   {Bird} S.,  2017, \mn@doi [\apj]
  {10.3847/1538-4357/835/2/175}, 835, 175

\bibitem[\protect\citeauthoryear{{Haardt} \& {Madau}}{{Haardt} \&
  {Madau}}{2001}]{Haardt_2001}
{Haardt} F.,  {Madau} P.,  2001, in Clusters of Galaxies and the High Redshift
  Universe Observed in X-rays, {Neumann}, D.~M. \& {Tran}, J.~T.~V. ed.,
  astro-ph/0106018.

\bibitem[\protect\citeauthoryear{{Haardt} \& {Madau}}{{Haardt} \&
  {Madau}}{2012}]{Haardt_2012ApJ}
{Haardt} F.,  {Madau} P.,  2012, \mn@doi [\apj] {10.1088/0004-637X/746/2/125},
  746, 125

\bibitem[\protect\citeauthoryear{{Hu}, {Kim}, {Cowie}, {Songaila}  \&
  {Rauch}}{{Hu} et~al.}{1995}]{Hu1995}
{Hu} E.~M.,  {Kim} T.-S.,  {Cowie} L.~L.,  {Songaila} A.,   {Rauch} M.,  1995,
  \mn@doi [\aj] {10.1086/117625}, 110, 1526

\bibitem[\protect\citeauthoryear{{Iapichino}, {Viel}  \& {Borgani}}{{Iapichino}
  et~al.}{2013}]{Iapichino2013}
{Iapichino} L.,  {Viel} M.,   {Borgani} S.,  2013, \mn@doi [\mnras]
  {10.1093/mnras/stt611}, 432, 2529

\bibitem[\protect\citeauthoryear{{Janknecht}, {Reimers}, {Lopez}  \&
  {Tytler}}{{Janknecht} et~al.}{2006}]{Janknecht_2006A&A}
{Janknecht} E.,  {Reimers} D.,  {Lopez} S.,   {Tytler} D.,  2006, \mn@doi
  [\aap] {10.1051/0004-6361:20065372}, 458, 427

\bibitem[\protect\citeauthoryear{{Khaire} \& {Srianand}}{{Khaire} \&
  {Srianand}}{2015}]{Khaire2015}
{Khaire} V.,  {Srianand} R.,  2015, \mn@doi [\mnras] {10.1093/mnrasl/slv060},
  451, L30

\bibitem[\protect\citeauthoryear{{Khaire}, {Srianand}, {Choudhury}  \&
  {Gaikwad}}{{Khaire} et~al.}{2016}]{Khaire_2016MNRAS}
{Khaire} V.,  {Srianand} R.,  {Choudhury} T.~R.,   {Gaikwad} P.,  2016, \mn@doi
  [\mnras] {10.1093/mnras/stw192}, 457, 4051

\bibitem[\protect\citeauthoryear{{Kim}, {Hu}, {Cowie}  \& {Songaila}}{{Kim}
  et~al.}{1997}]{Kim1997}
{Kim} T.-S.,  {Hu} E.~M.,  {Cowie} L.~L.,   {Songaila} A.,  1997, \mn@doi [\aj]
  {10.1086/118446}, 114, 1

\bibitem[\protect\citeauthoryear{{Kim}, {Cristiani}  \& {D'Odorico}}{{Kim}
  et~al.}{2001}]{Kim2001}
{Kim} T.-S.,  {Cristiani} S.,   {D'Odorico} S.,  2001, \mn@doi [\aap]
  {10.1051/0004-6361:20010650}, 373, 757

\bibitem[\protect\citeauthoryear{{Kim}, {Bolton}, {Viel}, {Haehnelt}  \&
  {Carswell}}{{Kim} et~al.}{2007}]{Kim_2007MNRAS}
{Kim} T.-S.,  {Bolton} J.~S.,  {Viel} M.,  {Haehnelt} M.~G.,   {Carswell}
  R.~F.,  2007, \mn@doi [\mnras] {10.1111/j.1365-2966.2007.12406.x}, 382, 1657

\bibitem[\protect\citeauthoryear{{Kim}, {Partl}, {Carswell}  \&
  {M{\"u}ller}}{{Kim} et~al.}{2013}]{Kim_2013A&A}
{Kim} T.-S.,  {Partl} A.~M.,  {Carswell} R.~F.,   {M{\"u}ller} V.,  2013,
  \mn@doi [\aap] {10.1051/0004-6361/201220042}, 552, A77

\bibitem[\protect\citeauthoryear{{Kirkman} \& {Tytler}}{{Kirkman} \&
  {Tytler}}{1997}]{Kirkman1997}
{Kirkman} D.,  {Tytler} D.,  1997, \mn@doi [\apj] {10.1086/304371}, 484, 672

\bibitem[\protect\citeauthoryear{{Kirkman}, {Tytler}, {Lubin}  \&
  {Charlton}}{{Kirkman} et~al.}{2007}]{Kirkman_2007MNRAS}
{Kirkman} D.,  {Tytler} D.,  {Lubin} D.,   {Charlton} J.,  2007, \mn@doi
  [\mnras] {10.1111/j.1365-2966.2007.11502.x}, 376, 1227

\bibitem[\protect\citeauthoryear{{Kollmeier} et~al.,}{{Kollmeier}
  et~al.}{2014}]{Kollmeier_2014ApJ}
{Kollmeier} J.~A.,  et~al., 2014, \mn@doi [\apjl]
  {10.1088/2041-8205/789/2/L32}, 789, L32

\bibitem[\protect\citeauthoryear{{Lehner}, {Savage}, {Richter}, {Sembach},
  {Tripp}  \& {Wakker}}{{Lehner} et~al.}{2007}]{Lehner_2007ApJ}
{Lehner} N.,  {Savage} B.~D.,  {Richter} P.,  {Sembach} K.~R.,  {Tripp} T.~M.,
   {Wakker} B.~P.,  2007, \mn@doi [\apj] {10.1086/511749}, 658, 680

\bibitem[\protect\citeauthoryear{{Lehner}, {Savage}, {Richter}, {Sembach},
  {Tripp}  \& {Wakker}}{{Lehner} et~al.}{2008}]{Lehner2008erratum}
{Lehner} N.,  {Savage} B.~D.,  {Richter} P.,  {Sembach} K.~R.,  {Tripp} T.~M.,
   {Wakker} B.~P.,  2008, \mn@doi [\apj] {10.1086/524357}, 674, 613

\bibitem[\protect\citeauthoryear{{Lewis}, {Challinor}  \& {Lasenby}}{{Lewis}
  et~al.}{2000}]{Lewis_2000ApJ}
{Lewis} A.,  {Challinor} A.,   {Lasenby} A.,  2000, \mn@doi [\apj]
  {10.1086/309179}, 538, 473

\bibitem[\protect\citeauthoryear{{Lu}, {Sargent}, {Womble}  \&
  {Takada-Hidai}}{{Lu} et~al.}{1996}]{Lu1996}
{Lu} L.,  {Sargent} W.~L.~W.,  {Womble} D.~S.,   {Takada-Hidai} M.,  1996,
  \mn@doi [\apj] {10.1086/526756}, 472, 509

\bibitem[\protect\citeauthoryear{{Madau} \& {Haardt}}{{Madau} \&
  {Haardt}}{2015}]{Madau2015}
{Madau} P.,  {Haardt} F.,  2015, \mn@doi [\apjl] {10.1088/2041-8205/813/1/L8},
  813, L8

\bibitem[\protect\citeauthoryear{{McQuinn}}{{McQuinn}}{2016}]{McQuinn2016}
{McQuinn} M.,  2016, \mn@doi [\araa] {10.1146/annurev-astro-082214-122355}, 54,
  313

\bibitem[\protect\citeauthoryear{{Meiksin}}{{Meiksin}}{2009}]{Meiksin2009}
{Meiksin} A.~A.,  2009, \mn@doi [Reviews of Modern Physics]
  {10.1103/RevModPhys.81.1405}, 81, 1405

\bibitem[\protect\citeauthoryear{{Meiksin}, {Bolton}  \& {Tittley}}{{Meiksin}
  et~al.}{2015}]{Meiksin2015}
{Meiksin} A.,  {Bolton} J.~S.,   {Tittley} E.~R.,  2015, \mn@doi [\mnras]
  {10.1093/mnras/stv1682}, 453, 899

\bibitem[\protect\citeauthoryear{{Oppenheimer} \& {Dav{\'e}}}{{Oppenheimer} \&
  {Dav{\'e}}}{2009}]{Oppenheimer2009}
{Oppenheimer} B.~D.,  {Dav{\'e}} R.,  2009, \mn@doi [\mnras]
  {10.1111/j.1365-2966.2009.14676.x}, 395, 1875

\bibitem[\protect\citeauthoryear{{Pachat}, {Narayanan}, {Muzahid}, {Khaire},
  {Srianand}, {Wakker}  \& {Savage}}{{Pachat} et~al.}{2016}]{Pachat_2016MNRAS}
{Pachat} S.,  {Narayanan} A.,  {Muzahid} S.,  {Khaire} V.,  {Srianand} R.,
  {Wakker} B.~P.,   {Savage} B.~D.,  2016, \mn@doi [\mnras]
  {10.1093/mnras/stw194}, 458, 733

\bibitem[\protect\citeauthoryear{{Paschos}, {Jena}, {Tytler}, {Kirkman}  \&
  {Norman}}{{Paschos} et~al.}{2009}]{Paschos_2009MNRAS}
{Paschos} P.,  {Jena} T.,  {Tytler} D.,  {Kirkman} D.,   {Norman} M.~L.,  2009,
  \mn@doi [\mnras] {10.1111/j.1365-2966.2009.15140.x}, 399, 1934

\bibitem[\protect\citeauthoryear{{Planck Collaboration} et~al.,}{{Planck
  Collaboration} et~al.}{2014}]{Planck_2014A&A}
{Planck Collaboration} et~al., 2014, \mn@doi [\aap]
  {10.1051/0004-6361/201423743}, 571, A31

\bibitem[\protect\citeauthoryear{{Puchwein} \& {Springel}}{{Puchwein} \&
  {Springel}}{2013}]{Puchwein_2013MNRAS}
{Puchwein} E.,  {Springel} V.,  2013, \mn@doi [\mnras] {10.1093/mnras/sts243},
  428, 2966

\bibitem[\protect\citeauthoryear{{Puchwein}, {Bolton}, {Haehnelt}, {Madau},
  {Becker}  \& {Haardt}}{{Puchwein} et~al.}{2015}]{Puchwein_2015MNRAS}
{Puchwein} E.,  {Bolton} J.~S.,  {Haehnelt} M.~G.,  {Madau} P.,  {Becker}
  G.~D.,   {Haardt} F.,  2015, \mn@doi [\mnras] {10.1093/mnras/stv773}, 450,
  4081

\bibitem[\protect\citeauthoryear{{Rahmati}, {Pawlik}, {Raicevic}  \&
  {Schaye}}{{Rahmati} et~al.}{2013}]{Rahmati_2013MNRAS}
{Rahmati} A.,  {Pawlik} A.~H.,  {Raicevic} M.,   {Schaye} J.,  2013, \mn@doi
  [\mnras] {10.1093/mnras/stt066}, 430, 2427

\bibitem[\protect\citeauthoryear{{Rauch} et~al.,}{{Rauch}
  et~al.}{1997}]{Rauch_1997ApJ}
{Rauch} M.,  et~al., 1997, \mn@doi [\apj] {10.1086/304765}, 489, 7

\bibitem[\protect\citeauthoryear{{Richter}, {Fang}  \& {Bryan}}{{Richter}
  et~al.}{2006}]{Richter_2006A&A}
{Richter} P.,  {Fang} T.,   {Bryan} G.~L.,  2006, \mn@doi [\aap]
  {10.1051/0004-6361:20054556}, 451, 767

\bibitem[\protect\citeauthoryear{{Ricotti}, {Gnedin}  \& {Shull}}{{Ricotti}
  et~al.}{2000}]{Ricotti_2000ApJ}
{Ricotti} M.,  {Gnedin} N.~Y.,   {Shull} J.~M.,  2000, \mn@doi [\apj]
  {10.1086/308733}, 534, 41

\bibitem[\protect\citeauthoryear{{Rollinde}, {Theuns}, {Schaye}, {P{\^a}ris}
  \& {Petitjean}}{{Rollinde} et~al.}{2013}]{Rollinde_2013MNRAS}
{Rollinde} E.,  {Theuns} T.,  {Schaye} J.,  {P{\^a}ris} I.,   {Petitjean} P.,
  2013, \mn@doi [\mnras] {10.1093/mnras/sts057}, 428, 540

\bibitem[\protect\citeauthoryear{{Savaglio} et~al.,}{{Savaglio}
  et~al.}{1999}]{Savaglio1999}
{Savaglio} S.,  et~al., 1999, \mn@doi [\apjl] {10.1086/311963}, 515, L5

\bibitem[\protect\citeauthoryear{{Schaye}}{{Schaye}}{2001}]{Schaye_2001ApJ}
{Schaye} J.,  2001, \mn@doi [\apj] {10.1086/322421}, 559, 507

\bibitem[\protect\citeauthoryear{{Schaye}, {Theuns}, {Leonard}  \&
  {Efstathiou}}{{Schaye} et~al.}{1999}]{Schaye_1999MNRAS}
{Schaye} J.,  {Theuns} T.,  {Leonard} A.,   {Efstathiou} G.,  1999, \mn@doi
  [\mnras] {10.1046/j.1365-8711.1999.02956.x}, 310, 57

\bibitem[\protect\citeauthoryear{{Schaye}, {Theuns}, {Rauch}, {Efstathiou}  \&
  {Sargent}}{{Schaye} et~al.}{2000}]{Schaye_2000MNRAS}
{Schaye} J.,  {Theuns} T.,  {Rauch} M.,  {Efstathiou} G.,   {Sargent} W.~L.~W.,
   2000, \mn@doi [\mnras] {10.1046/j.1365-8711.2000.03815.x}, 318, 817

\bibitem[\protect\citeauthoryear{{Schaye} et~al.,}{{Schaye}
  et~al.}{2010}]{Schaye2010}
{Schaye} J.,  et~al., 2010, \mn@doi [\mnras]
  {10.1111/j.1365-2966.2009.16029.x}, 402, 1536

\bibitem[\protect\citeauthoryear{{Scott}, {Bechtold}, {Morita}, {Dobrzycki}  \&
  {Kulkarni}}{{Scott} et~al.}{2002}]{Scott_2002ApJ}
{Scott} J.,  {Bechtold} J.,  {Morita} M.,  {Dobrzycki} A.,   {Kulkarni} V.~P.,
  2002, \mn@doi [\apj] {10.1086/339982}, 571, 665

\bibitem[\protect\citeauthoryear{{Sembach}, {Tripp}, {Savage}  \&
  {Richter}}{{Sembach} et~al.}{2004}]{Sembach2004}
{Sembach} K.~R.,  {Tripp} T.~M.,  {Savage} B.~D.,   {Richter} P.,  2004,
  \mn@doi [\apjs] {10.1086/425037}, 155, 351

\bibitem[\protect\citeauthoryear{{Shull}, {Smith}  \& {Danforth}}{{Shull}
  et~al.}{2012}]{Shull_2012ApJ}
{Shull} J.~M.,  {Smith} B.~D.,   {Danforth} C.~W.,  2012, \mn@doi [\apj]
  {10.1088/0004-637X/759/1/23}, 759, 23

\bibitem[\protect\citeauthoryear{{Shull}, {Danforth}  \& {Tilton}}{{Shull}
  et~al.}{2014}]{Shull_2014ApJ}
{Shull} J.~M.,  {Danforth} C.~W.,   {Tilton} E.~M.,  2014, \mn@doi [\apj]
  {10.1088/0004-637X/796/1/49}, 796, 49

\bibitem[\protect\citeauthoryear{{Shull}, {Moloney}, {Danforth}  \&
  {Tilton}}{{Shull} et~al.}{2015}]{Shull_2015ApJ}
{Shull} J.~M.,  {Moloney} J.,  {Danforth} C.~W.,   {Tilton} E.~M.,  2015,
  \mn@doi [\apj] {10.1088/0004-637X/811/1/3}, 811, 3

\bibitem[\protect\citeauthoryear{{Sijacki}, {Springel}, {Di Matteo}  \&
  {Hernquist}}{{Sijacki} et~al.}{2007}]{Sijacki2007}
{Sijacki} D.,  {Springel} V.,  {Di Matteo} T.,   {Hernquist} L.,  2007, \mn@doi
  [\mnras] {10.1111/j.1365-2966.2007.12153.x}, 380, 877

\bibitem[\protect\citeauthoryear{{Springel}}{{Springel}}{2005}]{Springel_2005M%
NRAS}
{Springel} V.,  2005, \mn@doi [\mnras] {10.1111/j.1365-2966.2005.09655.x}, 364,
  1105

\bibitem[\protect\citeauthoryear{{Springel} \& {Hernquist}}{{Springel} \&
  {Hernquist}}{2003}]{Springel_2003MNRAS}
{Springel} V.,  {Hernquist} L.,  2003, \mn@doi [\mnras]
  {10.1046/j.1365-8711.2003.06206.x}, 339, 289

\bibitem[\protect\citeauthoryear{{Springel} et~al.,}{{Springel}
  et~al.}{2005}]{Springel_2005Nat}
{Springel} V.,  et~al., 2005, \mn@doi [\nat] {10.1038/nature03597}, 435, 629

\bibitem[\protect\citeauthoryear{{Tepper-Garc{\'{\i}}a}}{{Tepper-Garc{\'{\i}}a%
}}{2006}]{Tepper-Garcia_2006MNRAS}
{Tepper-Garc{\'{\i}}a} T.,  2006, \mn@doi [\mnras]
  {10.1111/j.1365-2966.2006.10450.x}, 369, 2025

\bibitem[\protect\citeauthoryear{{Tepper-Garc{\'{\i}}a}, {Richter}, {Schaye},
  {Booth}, {Dalla Vecchia}  \& {Theuns}}{{Tepper-Garc{\'{\i}}a}
  et~al.}{2012}]{Tepper-Garcia_2012MNRAS}
{Tepper-Garc{\'{\i}}a} T.,  {Richter} P.,  {Schaye} J.,  {Booth} C.~M.,  {Dalla
  Vecchia} C.,   {Theuns} T.,  2012, \mn@doi [\mnras]
  {10.1111/j.1365-2966.2012.21545.x}, 425, 1640

\bibitem[\protect\citeauthoryear{{Theuns}, {Leonard}  \& {Efstathiou}}{{Theuns}
  et~al.}{1998a}]{Theuns1998_lowz}
{Theuns} T.,  {Leonard} A.,   {Efstathiou} G.,  1998a, \mn@doi [\mnras]
  {10.1046/j.1365-8711.1998.01740.x}, 297, L49

\bibitem[\protect\citeauthoryear{{Theuns}, {Leonard}, {Efstathiou}, {Pearce}
  \& {Thomas}}{{Theuns} et~al.}{1998b}]{Theuns_1998MNRAS}
{Theuns} T.,  {Leonard} A.,  {Efstathiou} G.,  {Pearce} F.~R.,   {Thomas}
  P.~A.,  1998b, \mn@doi [\mnras] {10.1046/j.1365-8711.1998.02040.x}, 301, 478

\bibitem[\protect\citeauthoryear{{Theuns}, {Viel}, {Kay}, {Schaye}, {Carswell}
  \& {Tzanavaris}}{{Theuns} et~al.}{2002}]{Theuns2002}
{Theuns} T.,  {Viel} M.,  {Kay} S.,  {Schaye} J.,  {Carswell} R.~F.,
  {Tzanavaris} P.,  2002, \mn@doi [\apjl] {10.1086/344521}, 578, L5

\bibitem[\protect\citeauthoryear{{Tilton}, {Danforth}, {Shull}  \&
  {Ross}}{{Tilton} et~al.}{2012}]{Tilton_2012ApJ}
{Tilton} E.~M.,  {Danforth} C.~W.,  {Shull} J.~M.,   {Ross} T.~L.,  2012,
  \mn@doi [\apj] {10.1088/0004-637X/759/2/112}, 759, 112

\bibitem[\protect\citeauthoryear{{Upton Sanderbeck}, {D'Aloisio}  \&
  {McQuinn}}{{Upton Sanderbeck} et~al.}{2016}]{Sanderbeck_2016MNRAS}
{Upton Sanderbeck} P.~R.,  {D'Aloisio} A.,   {McQuinn} M.~J.,  2016, \mn@doi
  [\mnras] {10.1093/mnras/stw1117}, 460, 1885

\bibitem[\protect\citeauthoryear{{Viel}, {Haehnelt}  \& {Springel}}{{Viel}
  et~al.}{2004}]{Viel2004}
{Viel} M.,  {Haehnelt} M.~G.,   {Springel} V.,  2004, \mn@doi [\mnras]
  {10.1111/j.1365-2966.2004.08224.x}, 354, 684

\bibitem[\protect\citeauthoryear{{Viel}, {Schaye}  \& {Booth}}{{Viel}
  et~al.}{2013}]{Viel_2013MNRAS}
{Viel} M.,  {Schaye} J.,   {Booth} C.~M.,  2013, \mn@doi [\mnras]
  {10.1093/mnras/sts465}, 429, 1734

\bibitem[\protect\citeauthoryear{{Viel}, {Haehnelt}, {Bolton}, {Kim},
  {Puchwein}, {Nasir}  \& {Wakker}}{{Viel} et~al.}{2017}]{Viel_2016}
{Viel} M.,  {Haehnelt} M.~G.,  {Bolton} J.~S.,  {Kim} T.-S.,  {Puchwein} E.,
  {Nasir} F.,   {Wakker} B.~P.,  2017, \mn@doi [\mnras]
  {10.1093/mnrasl/slx004}, 467, L86

\bibitem[\protect\citeauthoryear{{Vogelsberger} et~al.,}{{Vogelsberger}
  et~al.}{2014}]{Vogelsberger2014}
{Vogelsberger} M.,  et~al., 2014, \mn@doi [\mnras] {10.1093/mnras/stu1536},
  444, 1518

\bibitem[\protect\citeauthoryear{{Wakker}, {Hernandez}, {French}, {Kim},
  {Oppenheimer}  \& {Savage}}{{Wakker} et~al.}{2015}]{Wakker_2015ApJ}
{Wakker} B.~P.,  {Hernandez} A.~K.,  {French} D.~M.,  {Kim} T.-S.,
  {Oppenheimer} B.~D.,   {Savage} B.~D.,  2015, \mn@doi [\apj]
  {10.1088/0004-637X/814/1/40}, 814, 40

\bibitem[\protect\citeauthoryear{{Weymann} et~al.,}{{Weymann}
  et~al.}{1998}]{Weymann_1998ApJ}
{Weymann} R.~J.,  et~al., 1998, \mn@doi [\apj] {10.1086/306233}, 506, 1

\bibitem[\protect\citeauthoryear{{Williger}, {Heap}, {Weymann}, {Dav{\'e}},
  {Ellingson}, {Carswell}, {Tripp}  \& {Jenkins}}{{Williger}
  et~al.}{2006}]{Williger2006}
{Williger} G.~M.,  {Heap} S.~R.,  {Weymann} R.~J.,  {Dav{\'e}} R.,  {Ellingson}
  E.,  {Carswell} R.~F.,  {Tripp} T.~M.,   {Jenkins} E.~B.,  2006, \mn@doi
  [\apj] {10.1086/498127}, 636, 631

\bibitem[\protect\citeauthoryear{{Williger} et~al.,}{{Williger}
  et~al.}{2010}]{Williger_2010MNRAS}
{Williger} G.~M.,  et~al., 2010, \mn@doi [\mnras]
  {10.1111/j.1365-2966.2010.16519.x}, 405, 1736

\makeatother
\end{thebibliography}

\appendix
\section{numerical convergence}
\begin{figure*}
\includegraphics[width=.85\textwidth,trim={0.0cm 5.2cm 0.1cm 1.4cm},clip]{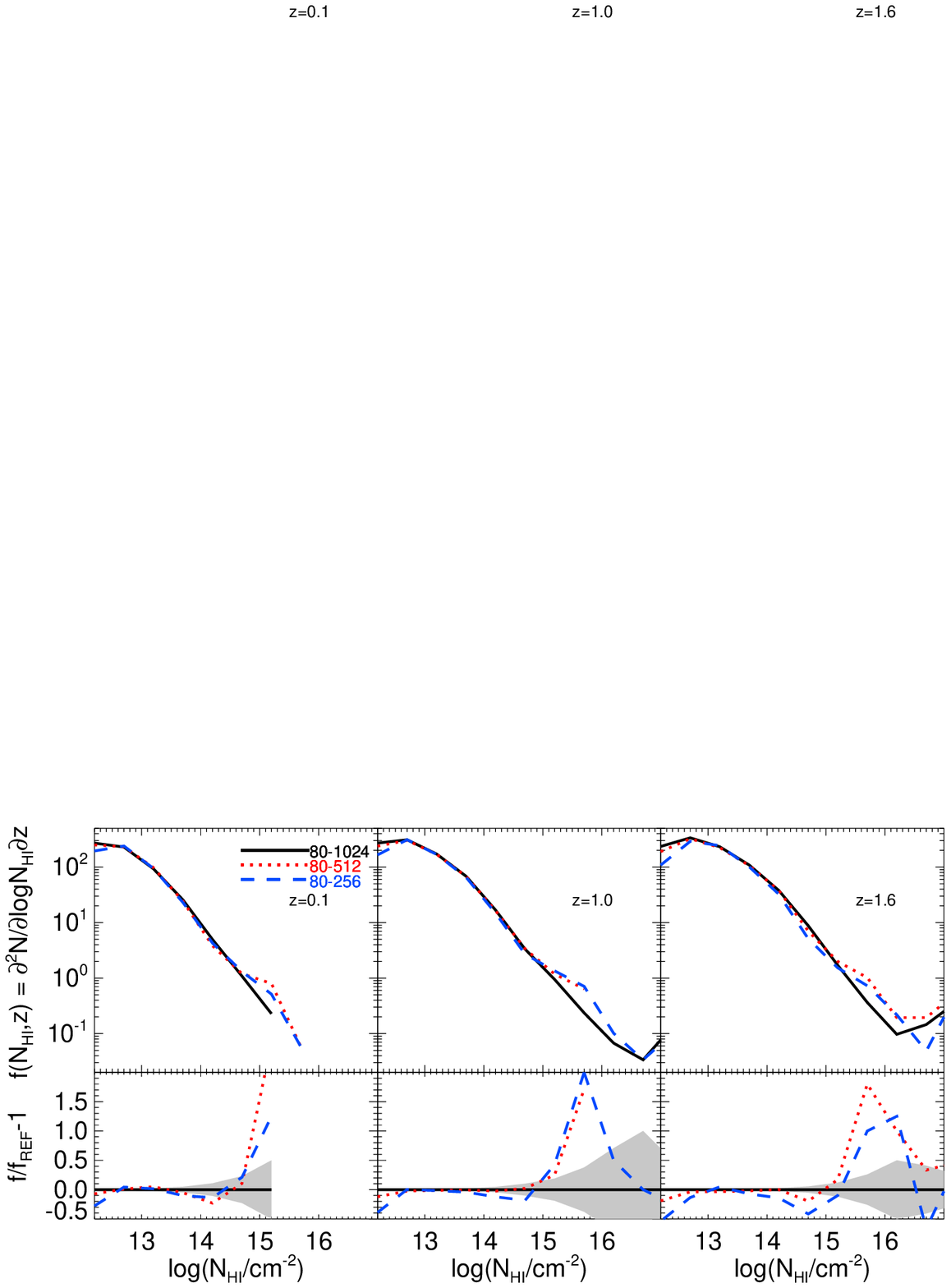}       
\vspace{-1.5cm}
\caption{The \Lya forest CDDF at $z=0.1,\,1,\,$ and $1.6$.
  Convergence with mass resolution for a fixed box size of $80$\mpc is
  displayed. Only column densities with a relative error less then 50
  per cent are used. }
\label{fig:CDDF_RRes} 
\end{figure*}

\begin{figure*}
\includegraphics[width=.85\textwidth,trim={0.0cm 5.2cm 0.1cm 1.4cm},clip]{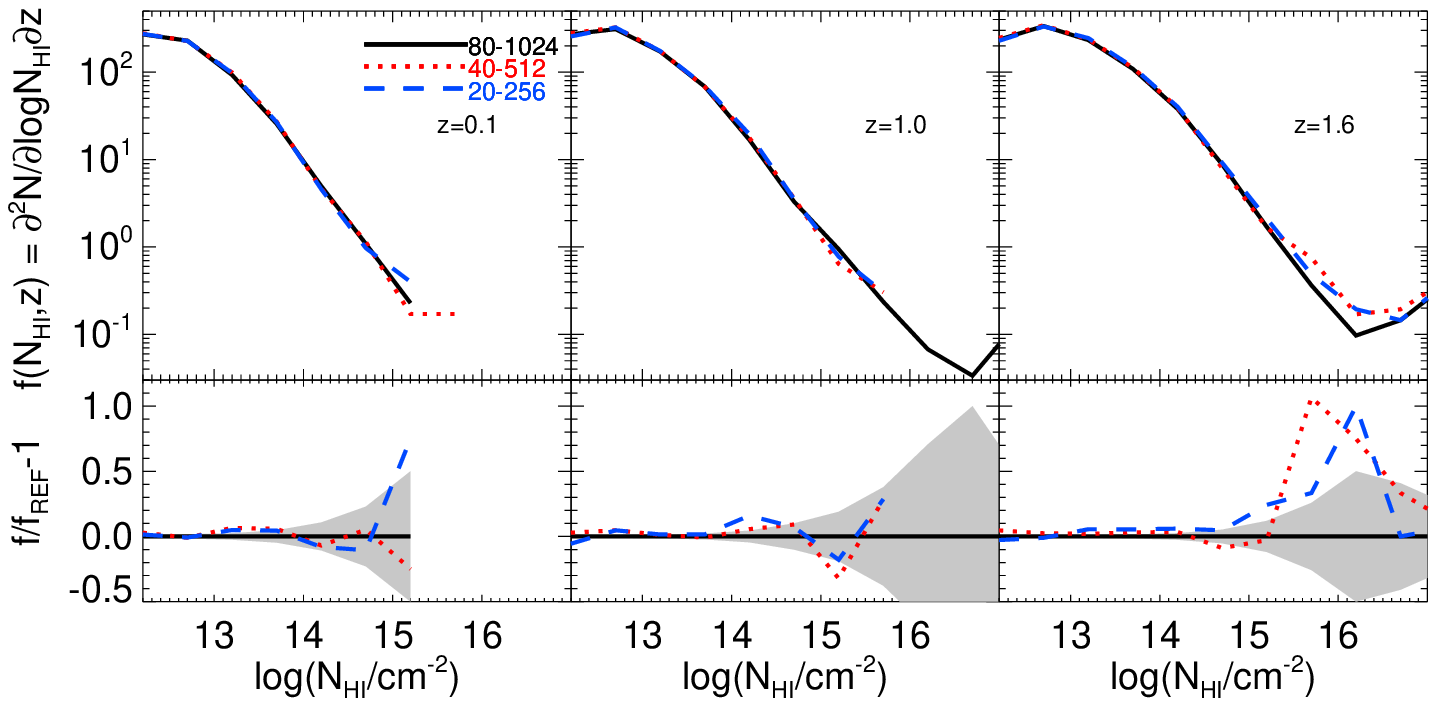}       
\vspace{-1.5cm}
\caption{As for Figure~\ref{fig:CDDF_RRes}, but now convergence with box
  size for a fixed mass resolution of $M_{\rm gas}=6.38\times10^6
  h^{-1} M_{\sun} $ is displayed.}
\label{fig:CDDF_RBox} 
\end{figure*}

\begin{figure*}
\includegraphics[width=.85\textwidth,trim={0.0cm 5.2cm 0.1cm 1.4cm},clip]{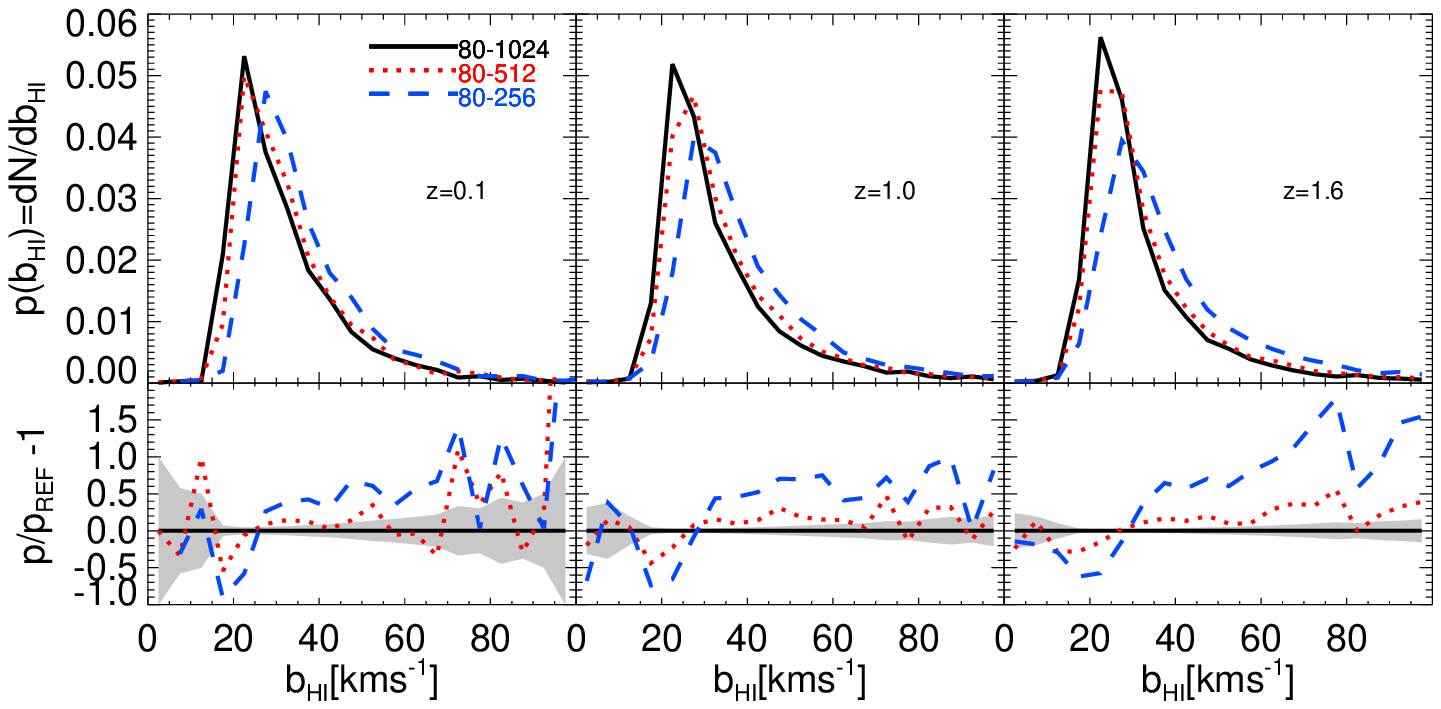}       
\vspace{-1.5cm}
\caption{The \Lya velocity width distribution at $z=0.1,\,1,\,$ and
  $1.6$.  Convergence with mass resolution for a fixed box size of
  $80$\mpc is displayed. Only lines with $N_{\rm
    HI}=10^{13}$--$10^{14}\rm\,cm^{-2}$ and a relative error less than
  50 per cent on the velocity widths are used. }
\label{fig:BDist_RRes} 
\end{figure*}

\begin{figure*}
\includegraphics[width=.85\textwidth,trim={0.0cm 5.2cm 0.1cm 1.4cm},clip]{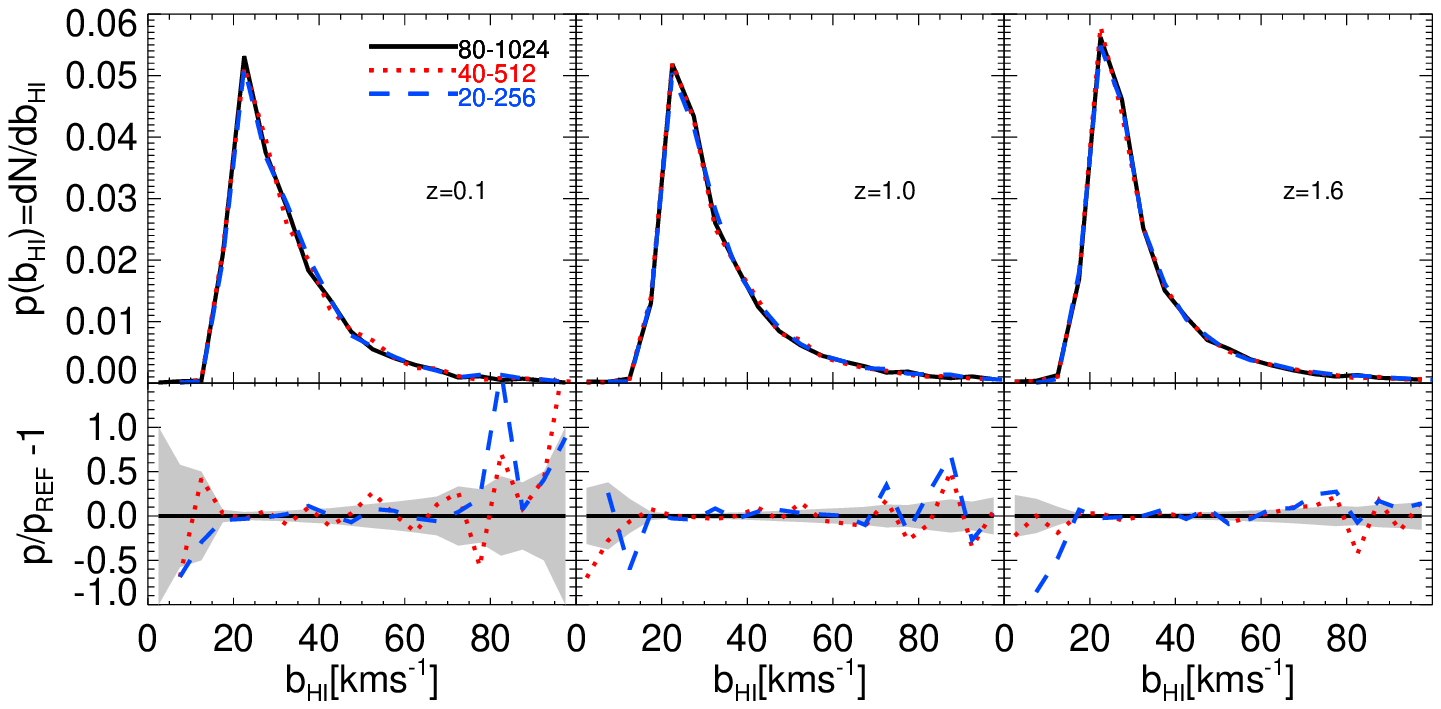}       
\vspace{-1.5cm}
\caption{As for Figure~\ref{fig:BDist_RRes}, but now convergence with box
  size for a fixed mass resolution of $M_{\rm gas}=6.38\times10^6
  h^{-1} M_{\sun} $ is displayed.}
\label{fig:BDist_RBox} 
\end{figure*}

We have performed convergence tests with mass resolution and box size
for the \Lya forest CDDF and velocity width distribution at the
redshifts considered in this paper.  These are displayed in
Figures~\ref{fig:CDDF_RRes}-\ref{fig:BDist_RBox}.  The CDDF is well
converged with mass resolution and box size within the range $
10^{12.7}\leq \ N_{\rm HI}/\rm{cm}^{-2} \leq 10^{14}$.  However, at
$N_{\rm HI}>10^{14}\rm\,cm^{-2}$ ($N_{\rm HI}<10^{12.7}\rm\,cm^{-2}$)
the number of lines are over (under) estimated at lower mass
resolution.  This suggests that numerical resolution is unlikely to be
the cause of the discrepancy between the COS observations and our
simulations at $N_{\rm HI}>10^{14}\rm\,cm^{-2}$.

In Figure~\ref{fig:BDist_RRes}, however, it is clear that mass
resolution significantly impacts on the velocity width distributions,
with line widths that are systematically overestimated at our fiducial
resolution of $M_{\rm gas}=5.1\times 10^{7}h^{-1}M_{\odot}$
(80-512). A mass resolution of at least $M_{\rm gas}=6.4\times
10^{6}h^{-1}M_{\odot}$ is required for an acceptable level of
convergence.  This means the discrepancy between the COS observations
of the velocity width distribution will be even larger for higher
  resolution simulations.  The \Lya line width distribution is
generally well converged with box size, as shown
Figure~\ref{fig:BDist_RBox}, with the exception of lines at $b_{\rm
  HI}<20$\kms in the $20h^{-1}\rm\,cMpc$ box.

Finally, Figures~\ref{fig:BDist_RRes}-\ref{fig:BDist_RBox} and
  Table~\ref{tab:conv} display convergence tests for the volume
  weighted IGM thermal state with box size and mass resolution.  The
  results are well converged with box size, but demonstrate the
  temperature of the IGM at mean density is overestimated by
  $10$--$20$ per cent at our fiducial resolution (80-512).  We have
  checked, however, that the \emph{mass weighted} temperature obtained
  from the gas particles is well converged with mass resolution,
  suggesting this difference originates from the larger smoothing
  lengths used in the lower resolution models.  Note also that
  $T(\bar{\Delta})$, shown in the middle panels of
  Figures~\ref{fig:BDist_RRes}-\ref{fig:BDist_RBox}, corresponds to
  $T(\bar{\Delta})=T_{0}\bar{\Delta}^{\gamma-1}$, and therefore
  reflects the convergence of both $T_{0}$ and $\gamma-1$.
  Additionally, we observe from Table~\ref{tab:conv} that as mass
  resolution is increased using the \textsc{QUICKLYA} model, a greater
  proportion by mass of the baryons are converted to collisionless
  star particles.  Note, however, this is already a significant
  overestimate and the results of the \textsc{QUICKLYA} model will be
  unreliable for gas at $\Delta>\Delta_{\rm th}$.

\begin{figure*}
    \centering
    \begin{minipage}{.33\textwidth}
        \centering
        \includegraphics[trim={0.5cm 0.3cm 0.5cm 1.0cm}, clip=true, width=\columnwidth]{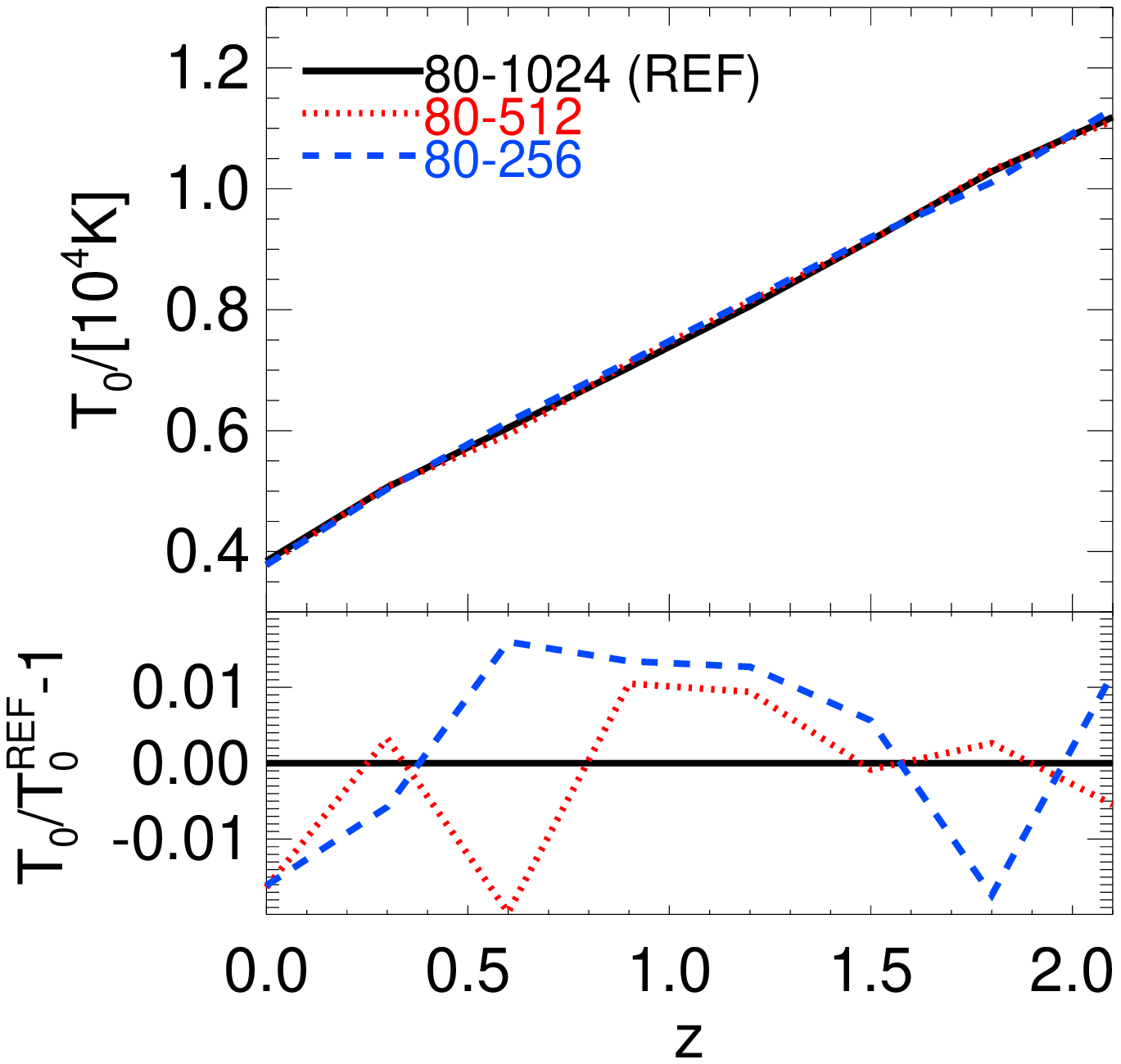}       
    \end{minipage}
    \begin{minipage}{.33\textwidth}
        \centering
        \includegraphics[trim={0.5cm 0.2cm 0.5cm 1.0cm}, clip=true, width=\columnwidth]{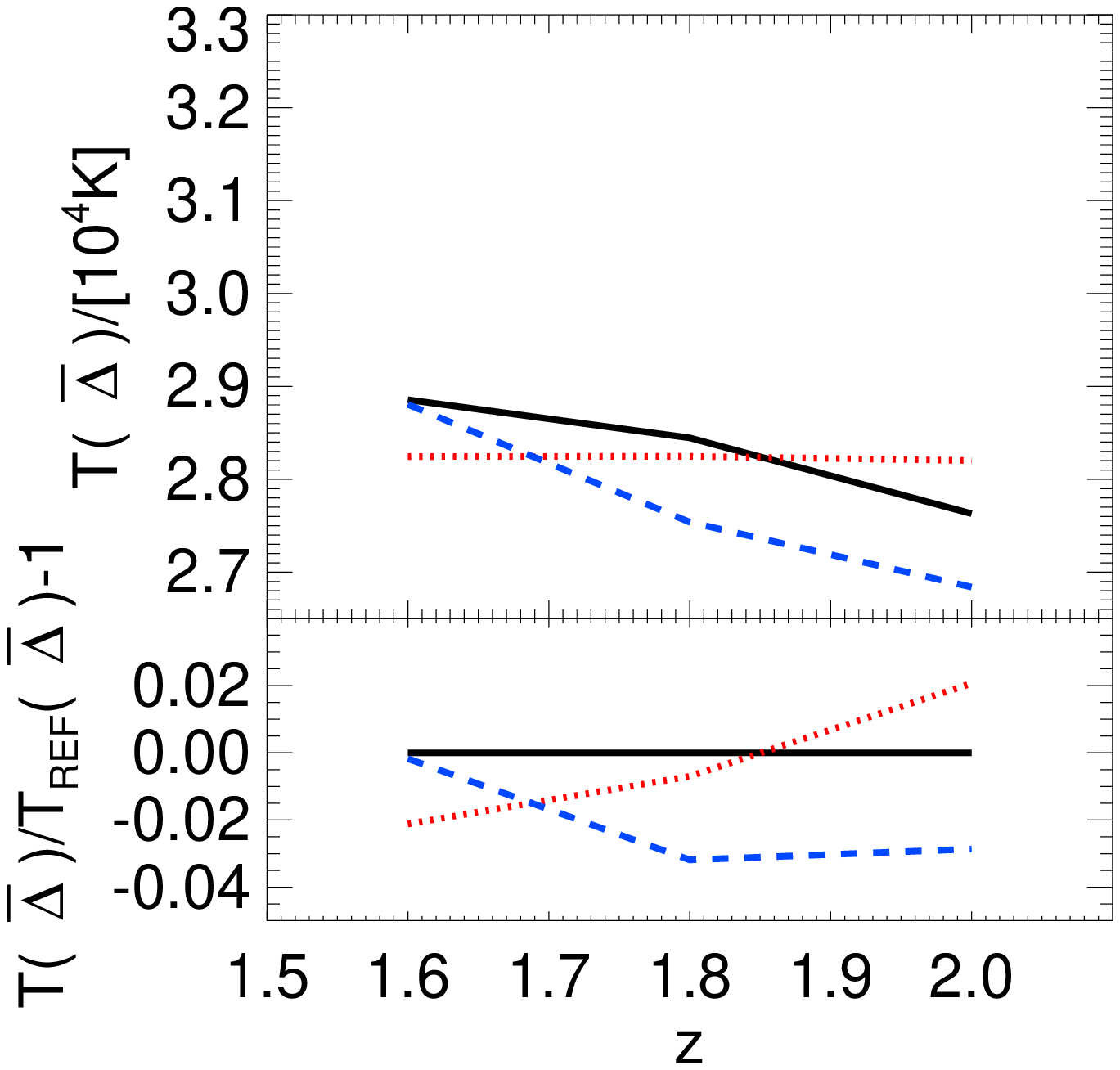}        
     \end{minipage}
     \begin{minipage}{.33\textwidth}
        \centering
        \includegraphics[trim={0.5cm 0.3cm 0.5cm 1.0cm}, clip=true, width=\columnwidth]{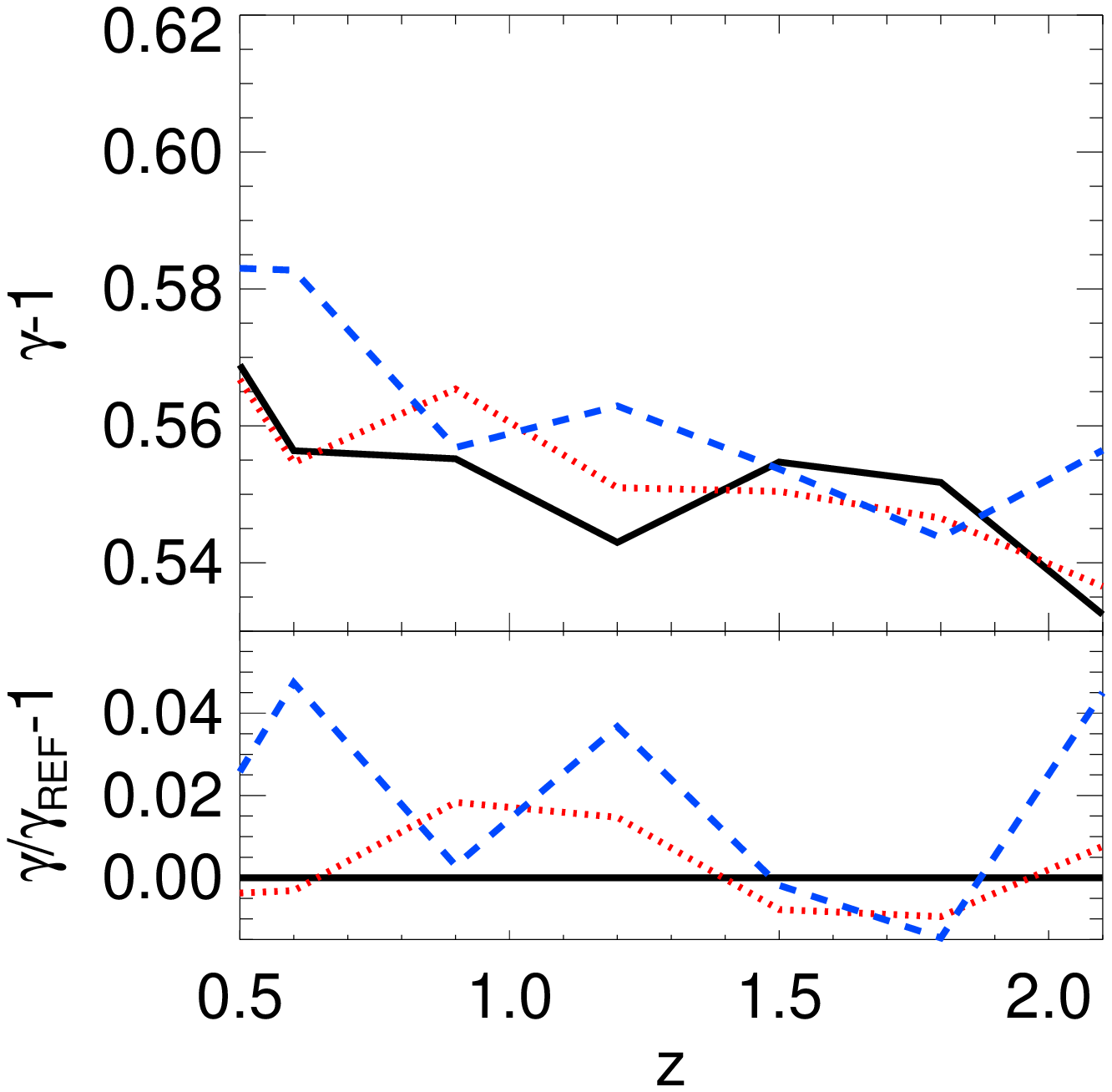}       
    \end{minipage}
\caption{The temperature at mean density $T_{0}$ (left), the
    temperature at the characteristic density, $T(\bar{\Delta})$,
    probed by the \Lya forest determined by \citet{Boera_2014}
    (middle) and the slope of the temperature-density relation
    $\gamma-1$ (right). Convergence with box size for a fixed mass
    resolution of $M_{\rm gas}=6.38\times10^6 h^{-1} M_{\sun} $ is
    displayed.}
\label{fig:TBoxRatio}
\end{figure*}

\begin{figure*}
    \centering
    \begin{minipage}{.33\textwidth}
        \centering
        \includegraphics[trim={0.5cm 0.3cm 0.5cm 1.0cm}, clip=true, width=\columnwidth]{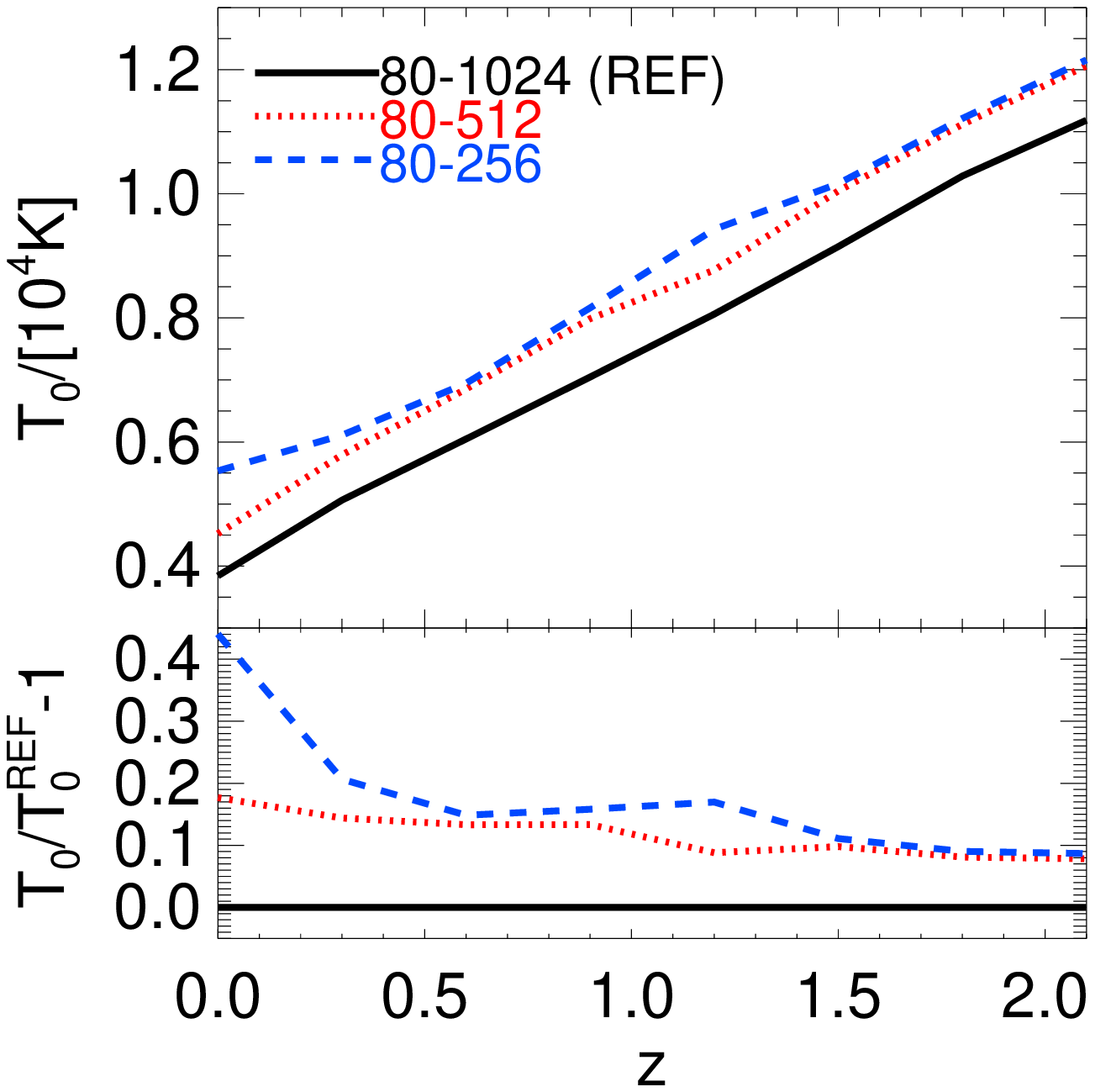}       
    \end{minipage}
    \begin{minipage}{.33\textwidth}
        \centering \includegraphics[trim={0.5cm 0.2cm 0.5cm 1.0cm},
          clip=true, width=\columnwidth]{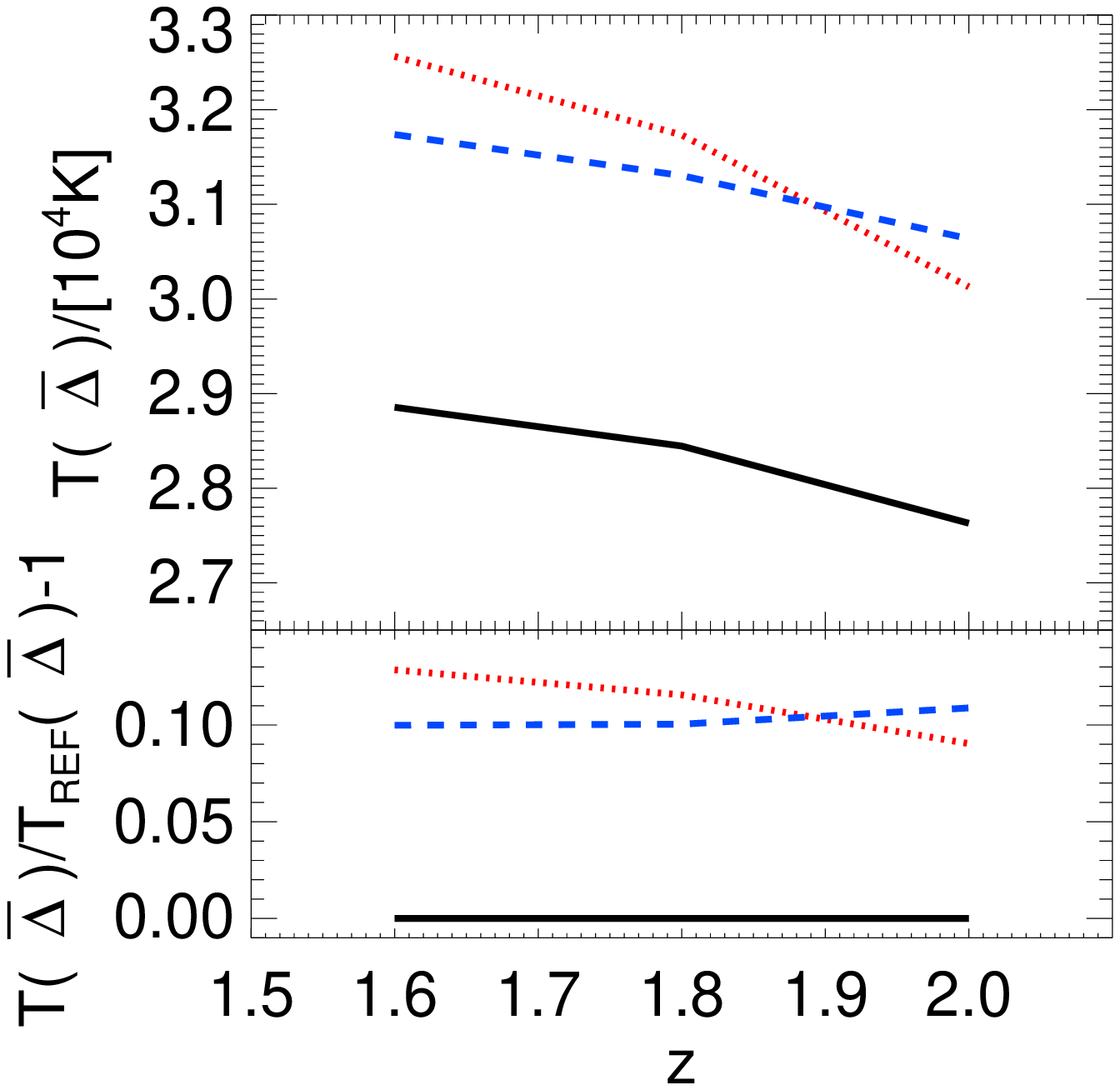}
     \end{minipage}
     \begin{minipage}{.33\textwidth}
        \centering
        \includegraphics[trim={0.5cm 0.3cm 0.5cm 1.0cm}, clip=true, width=\columnwidth]{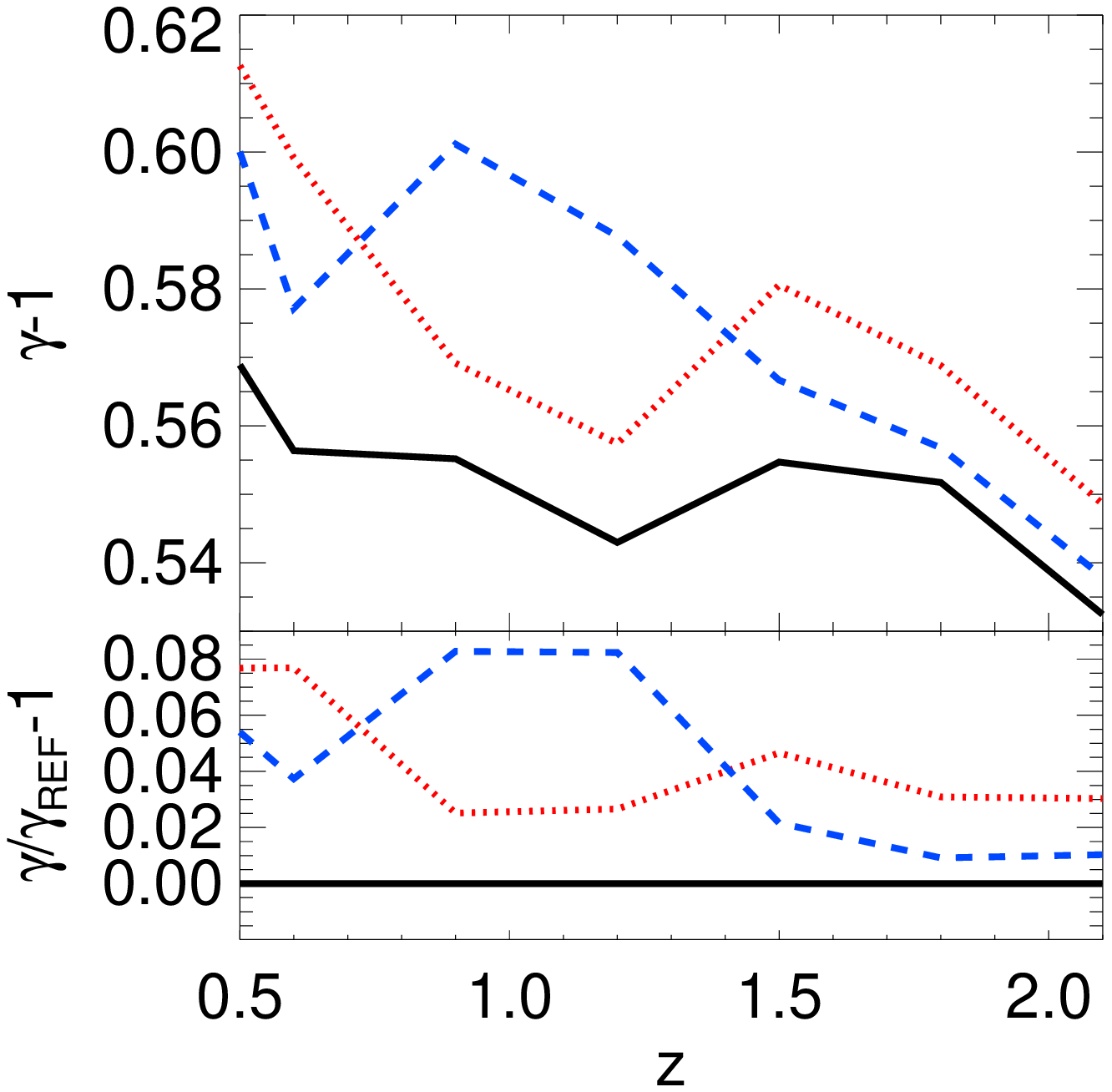}        \end{minipage}
\caption{As for Figure~\ref{fig:TBoxRatio}, but now convergence with
    mass resolution for a fixed box size of $80$\mpc is displayed.}
\label{fig:TResRatio}
\end{figure*}

\begin{table*}
\caption{As for Table \ref{tab:feedback}, but now including all
    remaining models using the \textsc{QUICKLYA} prescription for the
    removal of dense gas from the simulations. We repeat the entry for
    the 80-512 model here for ease of comparison. }\centering
\begin{tabular}{clcccccc}
\hline

Name             & Model & Diffuse & WHIM & Hot Halo & Condensed & Stars\\               
                 &       &$\Delta<\Delta_{\rm th},T<10^5$K&$\Delta<\Delta_{\rm th},T>10^5{\rm K}$&$\Delta >\Delta_{\rm th},T>10^5$K&$\Delta>\Delta_{\rm th},T<10^5$K\\
  \hline     
  $z=0.1$        & 80-1024         &35.0 &20.1 &15.3 &0.7 &28.8\\
                 & 80-512          &36.5 &21.3 &17.1 &0.9 &24.2\\  
                 & 80-256          &39.7 &22.6 &20.0 &1.2 &16.6\\  
                 & 40-512          &34.6 &20.5 &15.3 &0.8 &28.8\\
                 & 20-256          &35.3 &19.4 &15.2 &0.7 &29.3\\           

\hline           
 $z=1.0$         & 80-1024         &49.6 &14.5 &11.1 &1.6   &23.2\\
                 & 80-512          &51.7 &15.4 &12.6 &1.9   &18.4\\  
                 & 80-256          &55.4 &16.4 &14.9 &2.1   &11.2\\
                 & 40-512          &49.2 &15.0 &11.1 &1.5   &23.2\\
                 & 20-256          &49.7 &13.5 &11.5 &1.6   &23.7\\                           
       
\hline
 $z=1.6$         & 80-1024         &59.3 &10.7 &7.7  &2.3  &20.0\\
                 & 80-512          &61.8 &11.3 &8.9  &2.8  &15.1\\
                 & 80-256          &65.7 &12.3 &10.6 &2.9  &8.4 \\
                 & 40-512          &58.9 &11.1 &7.8  &2.2  &19.9\\
                 & 20-256          &59.2 &10.2 &7.9  & 2.3 &20.4\\

\hline
\end{tabular}
\label{tab:conv}
\end{table*}

\end{document}